\documentstyle[draft]{mn}

\part*{}

\title[Modelling protostellar infall]{Modelling 
submillimetre spectra of the protostellar infall candidates
NGC1333--IRAS2 and Serpens~SMM4}
\author[D. Ward-Thompson and H. D. Buckley]
{D. Ward-Thompson$^1$ and H. D. Buckley$^2$ \\
$^1$Dept of Physics \& Astronomy, Cardiff University, PO Box 913, Cardiff
CF2 3YJ \\
$^2$Institute for Astronomy, University of Edinburgh,
Blackford Hill, Edinburgh EH9 3HJ }

\date{Accepted 2001 July 1; Received 2000 December 13.} 
\pagerange{000--000} 

\def\LaTeX{L\kern-.36em\raise.3ex\hbox{a}\kern-.15em 
T\kern-.1667em\lower.7ex\hbox{E}\kern-.125emX} 

\newcommand{\msolar}{\mbox{\,$\rm M_{\odot}$}} 
\newcommand{\hcoft}{HCO$^+(J=4\rightarrow 3)$ }
\newcommand{\hcott}{HCO$^+(J=3\rightarrow 2)$ }
\newcommand{\htcoft}{H$^{13}$CO$^+(J=4\rightarrow 3)$ }
\newcommand{\htcott}{H$^{13}$CO$^+(J=3\rightarrow 2)$ }
\newcommand{\csff}{CS$(J=5\rightarrow 4)$ }
\newcommand{\csss}{CS$(J=7\rightarrow 6)$ }
\newcommand{\cott}{CO$(J=3\rightarrow 2)$ }

\newcommand{\ceott}{C$^{18}$O$(J=3\rightarrow 2)$ }
\newcommand{\ceoto}{C$^{18}$O$(J=2\rightarrow 1)$ }
\newcommand{\kms}{\,km\,s$^{-1}$}
\newcommand{\hco}{\mbox{HCO$^+$ }}
\newcommand{\htco}{\mbox{H$^{13}$CO$^+$ }}

\newcommand{\lsun}{\mbox{\,L$_{\odot}$}}

\begin{document} 
\label{firstpage} 

\maketitle 

\begin{abstract} 
We present a radiative transfer model, which is applicable to the study 
of submillimetre spectral line observations of protostellar envelopes. 
The model uses an exact, non-LTE, spherically symmetric radiative transfer 
`Stenholm' method, which numerically solves the radiative transfer problem 
by the process of `$\Lambda$-iteration'. We also present submillimetre 
spectral line data of the Class 0 protostars NGC1333--IRAS2 and Serpens~SMM4.
We model the data using the Stenholm code.

We examine the physical constraints which can be used to limit
the number and range of parameters used in protostellar envelope models,
and identify the turbulent velocity and tracer molecule abundance as the 
principle sources of uncertainty in the radiative transfer modelling. 
We explore the trends in the appearance of the predicted line profiles 
as key parameters in the models are varied, such as infall velocity
profile, turbulence and rotation.
The formation of the characteristic asymmetric double-peaked line profile in
infalling envelopes is discussed. {\em We find that the separation
of the two peaks of a typical infall profile is dependent not on the
evolutionary status of the collapsing protostar, but
on the turbulent velocity dispersion in the envelope}.
We also find that the line shapes can
be significantly altered by rotation.

Fits are found for the observed line profiles of IRAS2 and SMM4 using 
plausible infall model parameters. The density and velocity profiles
in our best fit models are inconsistent with a singular
isothermal sphere model (SIS), since for both
objects modelled, the infall velocities appear further advanced than a SIS
model would predict, given the density profile. We find better agreement 
with a form of collapse which assumes non-static initial conditions in
agreement with other recent findings.
We also find some evidence that the infall
velocities are retarded from free-fall towards the centre of the cloud,
probably by rotation, and that the envelope of SMM4 is rotationally 
flattened.

\end{abstract}
\begin{keywords}
stars: formation -- ISM: clouds -- radio continuum: ISM
\end{keywords}

\section{Introduction}

The process of star formation is currently not well understood. However,
the main protostellar collapse phases of low-mass stars 
($\sim$0.5--2M$_\odot$) have been identified observationally 
and labelled as Class 0 (Andr\'e, Ward-Thompson \& Barsony 1993), and 
Class I (Lada \& Wilking 1984; Lada 1987) protostars. These are
believed to represent the
phases during which the circumstellar envelope accretes
onto the central protostar and disk (e.g. Andr\'e 1994; Ward-Thompson 1996).
The final pre-main-sequence stages of Classes II \& III 
(Lada \& Wilking 1984; Lada 1987) correspond to the
Classical T Tauri (CTT) and Weak-line T Tauri (WTT) stages respectively
(Andr\'e \& Montmerle 1994). These protostellar stages are
beginning to be understood at least in
outline (for a review, see: Andr\'e, Ward-Thompson \& Barsony 2000).

Protostellar
infall has been reported in Class 0 sources by a number of authors
(e.g. Zhou et al. 1993; Ward-Thompson et al. 1996). However, the manner of
the collapse remains a matter for debate. The ideas of 
static initial conditions for
collapse (e.g. Shu 1977) have been disputed by many authors (e.g.
Foster \& Chevalier 1993; Whitworth et al. 1996). 
There is now a growing body of evidence which suggests
that the collapse occurs from non-static initial conditions,
and at a non-constant accretion rate which decreases with time (e.g.
Kenyon \& Hartmann 1995; Henriksen, Andr\'e \& Bontemps 1997;
Safier, McKee \& Stahler 1997; Whitworth \& Ward-Thompson 2001). 
Infall has even been detected 
in some starless cores, such as L1544 (Tafalla et al. 1999).

The densities and 
temperatures in the gas envelopes surrounding the youngest protostars are 
favourable for exciting a number of rotational molecular transitions, 
observable in the submillimetre waveband. The line profiles of these 
transitions contain information about both the physical state and dynamics 
of the envelope gas, and may potentially be used to test theoretical models 
of star formation, as many workers have shown (e.g. Bernes 1979; Rybicki \& 
Hummer 1991; Choi et al. 1995; Juvela 1997; Park \& Hong 1998; Hogerheijde 
\& van der Tak 2000).

In this paper observations are presented of the protostellar candidates
NGC1333--IRAS2 and Serpens SMM4, in transitions of 
HCO$^{+}$, H$^{13}$CO$^{+}$, CS, CO, $^{13}$CO and 
C$^{18}$O. The HCO$^{+}$ and CS transitions preferentially trace high density 
gas, whereas CO traces a much wider range of gas densities. 
The paper is laid out as follows: Section 2 introduces the 
$\Lambda$-iteration method of numerical radiative transfer; Section 3
describes our approach to the modelling; Section 4 explores the
sensitivity of the model to the various free parameters; Section 5
describes our observations and data reduction; Sections 6 \& 7 present
the results of our observations for NGC1333--IRAS2 and Serpens~SMM4
respectively; Section 8 compares the observations with the model 
predictions and finds the best fits to the data; Section 9 presents a brief
summary of our main findings. The non-expert reader may prefer to read
the second half of the paper first, starting from Section 5.

\section{Numerical Radiative Transfer}

Consider a cloud of gas with a specified distribution of density,
kinetic temperature and composition, which may have internal turbulent 
and systematic motions. Let any radiation sources not forming part of
the cloud itself also be specified. For each molecular species, there
exists a steady state solution for the distribution of rotational
energy level populations as a function of position in the cloud. In
the modelling of this paper we numerically calculate this distribution 
for the idealised case of a spherically symmetric model cloud, and
predict the observed line profiles. This is a complex
problem, due to the fact that well separated parts of the cloud 
can interact radiatively with each other. The $\Lambda$-iteration
method, described below, is conceptually one of the 
simplest techniques for solving this kind of problem. 

\begin{figure*}
\setlength{\unitlength}{1mm}
\begin{picture}(80,90)
\includegraphics{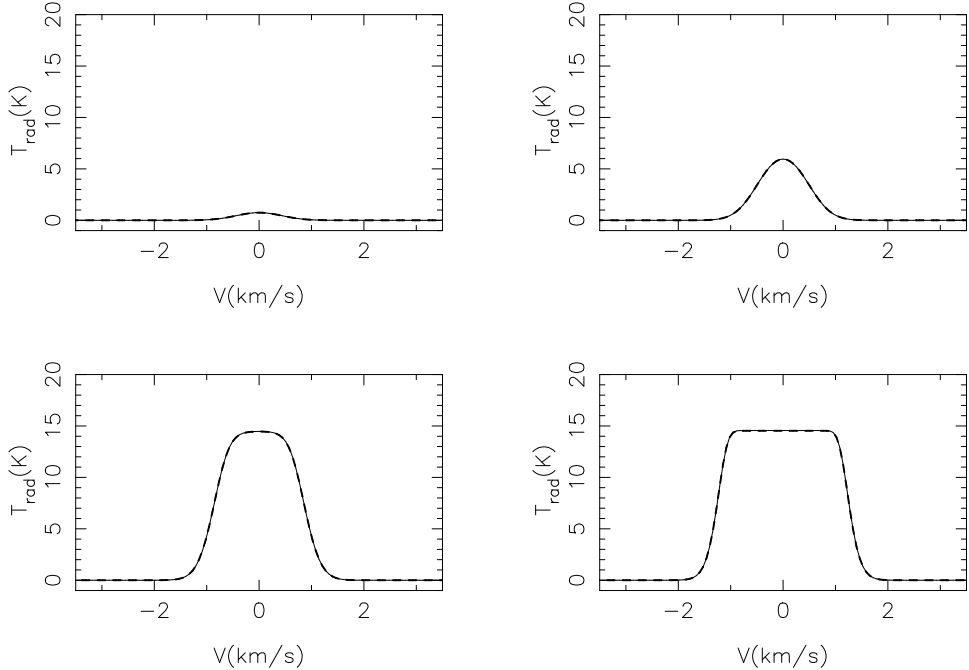}
\end{picture}
\caption{Comparison of analytical and numerical predictions of the 
CS(J$=5\rightarrow 4$) spectra from four uniform LTE models with 
progressively increasing CS abundance. The solid and dashed lines 
show the numerical and analytical
predictions respecitively. Each model has a radius of 10,000au, a uniform 
kinetic temperature of 20K, a uniform hydrogen number density of 
$10^{10}$cm$^{-3}$, 
and a uniform FWHM velocity width of 1.0km\,s$^{-1}$. The uniform 
CS number density
in the first model is $10^{-5}$cm$^{-3}$, and increases by factors of 10
in successive models, to
$10^{-2}$cm$^{-3}$ in the last model. The microwave background intensity has
been subtracted from both sets of spectra.} 
\label{LTE}
\end{figure*} 

\subsection{The $\Lambda$-iteration method}

The method is begun by choosing an initial 
radiation field in a more or less arbitrary manner. From this
a `false' run of level populations may
be calculated, by assuming the validity of the steady state rate 
equations. Radiative transitions between these level populations will 
generally produce a radiation field which departs from the one originally
assumed. If this radiation field is used to calculate a new set of
level populations, and the procedure is repeated a sufficiently large number
of times, the radiation field and level populations should
eventually converge on a mutually consistent
solution. $\Lambda$-iteration is a kind of diffusion process, where 
imbalances in the radiation field are smoothed out over a length-scale 
corresponding to approximately one optical depth at each iteration step. 

A number of radiative transfer models have been published (e.g.
Rawlings et al. 1992; Zhou 1992; Walker et al. 1994). However,
the modelling in this paper was carried out using a modified version of 
the Stenholm $\Lambda$-iteration code developed by Stenholm and
subsequently expanded by Little and co-workers at the University of Kent 
(Stenholm 1977; Matthews 1986; Heaton et al. 1993). 
The code uses the above method to solve the spectral line radiative
transfer problem for the rotational transitions of linear molecules in
a spherically symmetric model cloud. Radial profiles
of systematic velocity, temperature, density, tracer molecule 
abundance and micro-turbulent velocity dispersion may be specified.

The model cloud is discretised using a number of spherical shells,
and the level populations in each shell are determined 
in the iteration process from the calculated mean 
radiation intensity in the co-moving 
shell frame. Once the level 
populations have converged, a calculation is made of
a simulated spectral line observation on the cloud. This part of the code
carries out straightforward numerical integrations of the equation of 
transfer along parallel lines of sight through the cloud to find the
emergent intensities for a grid of impact parameters, 
which are then weighted according to a two-dimensional gaussian 
beam profile function. 

\subsection{Testing for convergence}

It is important to find a reliable 
test for the convergence of the model. One possible
condition is that the maximum fractional change $\eta (i,{\rm J})_N$, 
of the level population $n(i,{\rm J})$ between 
iterations $N-1$ and $N$, is less than some specified value 
$\alpha$:

\begin{equation}
\eta (i,{\rm J})_N = 
\frac{n(i,{\rm J})_N-n(i,{\rm J})_{N\!-\!1}}{n(i,{\rm J})_N} < \alpha,
\end{equation}

\noindent
for all $i$ and J, 
where the subscripts specify the iteration numbers. In many cases,
the value of $\eta(i,{\rm J})$ 
is a good estimate of the absolute fractional error in
the corresponding level population, in which case convergence is ensured by
setting a suitably low value of $\alpha$ (a value of $0.02$ or less is
usually found to be sufficient). However, when the rate of 
convergence is particularly slow, this 
test may give a misleading result,
since the fractional change between two successive 
iterations may then significantly 
underestimate the total change produced over a large number of subsequent
iterations. The rate
of decrease of $\eta$ should somehow be taken into account. 

Dickel \& Auer (1994) use the following procedure to
estimate the total fractional error $\eta_{\rm tot}(i,{\rm J})_N$ of the
level population $n(i,{\rm J})$ after the $N$'th iteration: 

\begin{equation}
\eta_{\rm tot}(i,{\rm J})_N\simeq \left|\frac{\eta (i,{\rm J})_N}{[ \eta
(i,{\rm J})_N/\eta (i,{\rm J})_{N\!-\!1}]-1}\right|.
\end{equation} 

\noindent
The convergence criterion is then $\eta_{\rm tot}(i,{\rm J})< \alpha$ for
all $i$ and J, for some specified $\alpha$ much less than 1.
Note that if $\eta(i,{\rm J})_N\simeq \eta(i,{\rm J})_{N-1}$ then 
$\eta_{\rm tot}(i,{\rm J})_N$
is much larger than $\eta(i,{\rm J})_N$ -- i.e. the total error is 
much larger than
the fractional change between successive iterations.

The rate of convergence is fastest when the optical depth of the cloud 
is small, or when the hydrogen number density is much greater than the 
critical 
thermalisation density. When the optical 
depth is small, radiation can travel across the whole cloud in one iteration, 
allowing the level populations in different parts of the cloud to adjust to 
each other
quickly. When the density is close to, or above, the critical density for the 
principal
thermally accessible transitions of the molecule, the level populations are 
determined
mainly by molecular collisions, and are insensitive to changes in the 
radiation field.

For optically thick regions where the density is below the 
critical density, the convergence rate is generally much 
slower. The low density results
in low collision rates, and the level populations therefore 
depend more sensitively on 
the radiation field, while the high optical depth means 
that the radiation field 
will take many iterations to adjust to the conditions in distant parts 
of the cloud. The code converges most rapidly near to LTE and 
when optically thin.

We have made several alterations to the code since it was made available 
to us. These have included changes to the internal structure of the code 
by, for example, improving the shell subdivision algorithm and 
incorporating rotation 
into the line profile calculation.
However, the former did not affect the conclusions, it merely gave the code more resolution.
Some of the more significant changes made are now discussed in detail.

\subsection{The Rybicki approximation}

An important modification made to the radiative transfer calculation
was the removal of the approximation of core saturation
used in the original code, known as the
Rybicki approximation (Rybicki 1972; Stenholm 1977; Rybicki 1984), which may 
be summarised as follows: suppose that at some point in the cloud the
following relation for the optical depth $\tau_\nu$ at frequency $\nu$
is satisfied for all directions {\bf r}$_0$ and solid angles 
${\bf \hat{k}}$:

\begin{equation}
\tau_{\nu}({\bf r}_0,z_b,\nu,{\bf \hat{k}}) > \gamma, \label{Tau}
\end{equation}

\noindent
where $z_b$ is the distance to the boundary of the model and $\gamma$ is some 
prescribed constant. The intensity I in all directions at that point
is defined in the Rybicki approximation to be equal to the local source 
function S, i.e:

\begin{equation}
I_{\nu}({\bf r}_0,\nu) = S_{\nu}({\bf r}_0,\nu). \label{S}
\end{equation}

This equation is only applied for the particular values
of $\nu$ for which $\tau_\nu >\gamma$.
For increasing values of $\gamma$, the
volume of the cloud to which the approximation applies 
becomes smaller, and disappears
altogether when $\gamma$ increases beyond the maximum 
optical depth of the cloud. For the problems we 
consider, 
the source function cannot be assumed to be uniform over distances 
corresponding to a few optical depths, and the Rybicki approximation 
cannot be used. 
We therefore do not use the Rybicki approximation in the modelling 
in this paper.

\begin{figure*}
\setlength{\unitlength}{1mm}
\begin{picture}(80,180)
\includegraphics{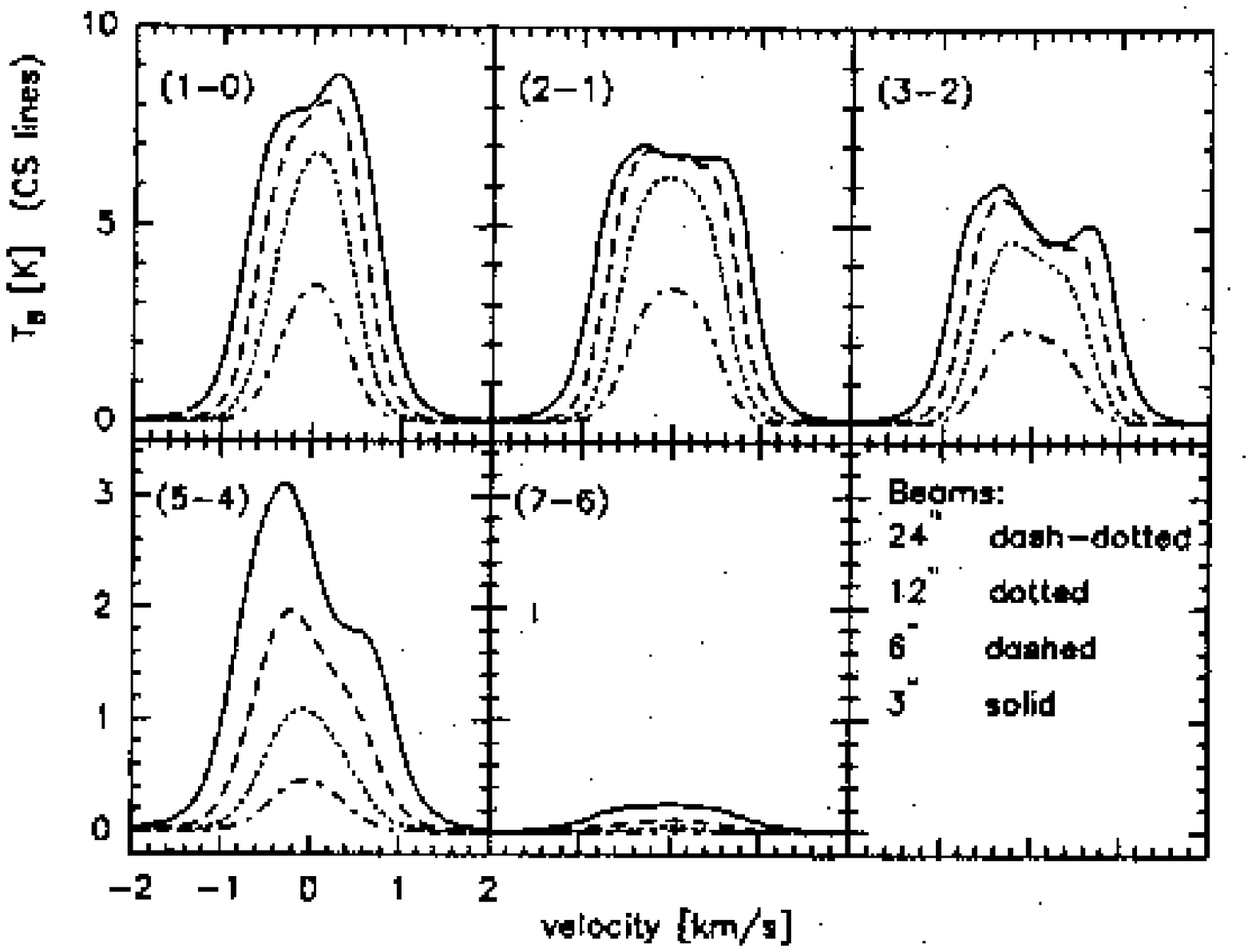}
\includegraphics{fig2b.eps}
\end{picture}
\caption{Comparison of the CS line profile calculations from figure 13 of
Kr\"ugel \& Chini (1994) -- top -- with those produced by Stenholm for the 
same model cloud -- lower. For each transition, 
line profiles are plotted for a number of different FWHM beam sizes, 
as indicated in the
bottom right hand panel of the top diagram. The model cloud has an 
outer radius of 
6684\,au and is assumed to lie at a distance of 450\,pc. The adopted 
hydrogen number
density and velocity profiles are of the form 
$n_{\rm H_2}(r)\propto\! r^{-3/2}$ and v$(r)\propto\! r^{-1/2}$ 
respectively, with $n_{\rm H_2}=1.1\times 10^5$ and 
v=$-$0.237\,km\,s$^{-1}$
at the outer edge of the cloud. The kinetic temperature is set 
to 10K up to the
half-point radius, rising linearly to a value of 20K at the outer
edge of the cloud. 
The microturbulent velocity width and the CS abundance are 
assumed to be constant
throughout the cloud, with values of 0.5\,km\,s$^{-1}$ (FWHM) 
and 2$\times 10^{-9}$ respectively.}
\label{krugel}
\end{figure*}

\subsection{Solid body rotation}

We have incorporated rotation into the code, in the form of
`solid body' rotation -- i.e. the angular velocity of all points in the 
cloud about some fixed axis is a constant, but systematic radial motions in 
the rotating 
frame are still allowed. The radiative transfer 
effects of solid body rotation were
fully accounted for by changes to the section of the code which
calculates the 
emergent line profile from the cloud, after the $\Lambda$-iteration 
solution has
been found. No change needs to be made to the $\Lambda$-iteration code 
itself, 
since the solution for the level populations in the cloud does not change 
when solid body rotation is introduced, as is now explained.

The radiation field in a frame co-moving with the gas
at a given point in the model can only be affected by a 
superimposed velocity field 
if the latter introduces changes in the relative line of sight
velocities of other 
parts of the cloud (ignoring relativistic effects of order $v^2/c^2$ 
or less). Solid body rotation does not introduce 
changes in relative line of sight velocities from one part 
of the cloud to another. 
For example, consider a non-rotating cloud which has systematic 
radial velocities 
${\bf {\rm v}_r}({\bf r_1})$ and ${\bf {\rm v}_r}({\bf r_2})$ at the 
positions ${\bf r_1}$ and ${\bf r_2}$ respectively. 
Writing $r_{12}=\mid\!{\bf r_2}\!-\!{\bf r_1}\!\mid$ 
for the distance separating
the two points, the component of relative 
velocity, ${\rm v}_{\rm los}$, along the line of 
sight between ${\bf r_1}$ and
${\bf r_2}$ is given by:

\begin{equation}
{\rm v}_{\rm los} = 
({\bf r_2}\!-\!{\bf r_1})\cdot 
[{\bf {\rm v}_r}({\bf r_2})\!-\!{\bf 
{\rm v}_r}({\bf r_1})]/r_{12}. \label{vlos1}
\end{equation}

Now assume the cloud is made to rotate in solid-body fashion with angular 
velocity ${\bf \Omega}$. Let the new systematic velocities at positions 
${\bf r_1}$ and ${\bf r_2}$ be ${\bf v}({\bf r_1})$ and 
${\bf v}({\bf r_2})$, where:

\begin{equation}
{\bf v}({\bf r_1}) = {\bf {\rm v}_r}({\bf r_1}) + 
{\bf \Omega} \times {\bf r_1},
\end{equation}

\begin{equation}
{\bf v}({\bf r_2}) = {\bf {\rm v}_r}({\bf r_2}) + 
{\bf \Omega} \times {\bf r_2}.
\end{equation}

Substituting into equation 5, the terms in ${\bf \Omega}$
disappear, and the final expression is unchanged.
This implies that the solid body rotation
has no effect on the relative line of sight velocities between any two
points in the cloud, and hence the radiation field in the co-moving
frame at each point in the cloud is not affected by solid body
rotation. This result only applies to the radiation emitted 
by the cloud itself. Radiation entering the model from outside will be 
Doppler shifted in the gas frame by the rotation, but the effect of
this is completely negligible for realistic external radiation fields.

However, solid body rotation does make a difference to the 
prediction for the observed
line profile, since the rotation does introduce changes in 
the line of sight velocities 
of different parts of the cloud relative to an observer 
outside the cloud.
If ${\bf \hat{z}}$ is the unit vector parallel to the line of 
sight from the observer
to the cloud, and ${\rm v}_z({\bf r})$ is the line of sight 
velocity (relative to the 
observer) of the gas at the position vector ${\bf r}$ (relative 
to the centre of the 
cloud), then:

\begin{equation}
{\rm v}_z({\bf r}) = {\bf \hat{z}}\cdot{\bf {\rm v}_r}({\bf r}) 
+ {\bf \hat{z}}\cdot({\bf \Omega}\times {\bf r}) = 
{\rm v}_{r,z}({\bf r}) + y\Omega_x-x\Omega_y,\label{vrot}
\end{equation}

\noindent
where ${\rm v}_{r,z}({\bf r})$ is the component along the line of
sight of the radial component of the gas velocity at ${\bf r}$, and
$x$ and $y$ are, as before, spatial
co-ordinates in the plane perpendicular to the line of sight at the
distance of the cloud. $\Omega_x$ and $\Omega_y$ are the components
of the angular velocity vector in this plane. 
The component of angular velocity parallel to the line of sight, 
$\Omega_z$, 
does not appear in equation 8 and has no effect on the observed
spectrum from the cloud.

\subsection{LTE comparison}

Analytical solutions of the equation of transfer are in general 
unobtainable for all but the simplest of problems. However, in 
the limit that the 
level populations are completely thermalised by collisions -- i.e. in
local thermodynamic equilibrium (LTE) --
the spectral line calculations from the numerical model may be compared
directly with analytical predictions.

In LTE, the excitation temperature for each transition approaches the 
kinetic temperature $T_{kin}$
of the gas, and the source function for the transition 
J$\rightarrow {\rm J}\!-\!1$ 
becomes $S_{\nu}({\bf r})=B_{\nu}[\nu_{\rm J},T_{\rm kin}(\bf r)]$.
If $T_{\rm kin}$ is constant over the cloud, then the solution
to the equation of transfer is:

\begin{equation}
I_{\nu}({\rm v}_z,{\rm J}) = 
I_{\nu}^{\rm ext}({\rm J})e^{-\tau_{\nu}({\rm v}_z)} + 
B_{\nu}(\nu_{\rm J},T_{\rm kin})\left(1\!-\! 
e^{-\tau_{\nu}({\rm v}_z)}\right), \label{soltcont2}
\end{equation}

\noindent
where $\tau_{\nu}({\rm v}_z)$ is the optical depth of 
the cloud at the frequency
corresponding to the velocity bin ${\rm v}_z$. For 
simplicity, we consider the intensity
emerging from a uniform density isothermal cloud layer of thickness $L$, 
with no
systematic motions, and with a constant velocity dispersion 
($\sigma_{\rm v}$) 
and total tracer molecule number density ($n_{\rm tot}$). 
The optical depth through the cloud layer may then be written as:

\begin{equation}
\tau_{\nu}({\rm v}_z) = \kappa (\nu)L
\end{equation}

\begin{equation}
=> \tau_\nu = \frac{h\nu_{\rm J} L}
{4\pi}[{\rm B}_{lu}({\rm J})n({\rm J}-1)-
{\rm B}_{ul}({\rm J})n({\rm J})]\phi({\rm v}_z)
\end{equation}

\begin{equation}
=> \tau_\nu = \frac{h\nu_{\rm J} L}
{4\pi}{\rm B}_{ul}n({\rm J}) 
[\exp (h\nu_{\rm J}/kT_{\rm kin})-1]\phi({\rm v}_z),
\end{equation}

\noindent
where $n({\rm J})$ and 
$n({\rm J}-1)$ are the number densities of tracer molecules in the upper and
lower level of the transition, and other symbols take their usual meanings.
In the last equation we have used the fact that in LTE:

\begin{equation}
\frac{n({\rm J}-1)}{n({\rm J})} = 
\frac{g({\rm J}-1)}{g({\rm J})}\exp (h\nu_{\rm J}/kT_{\rm kin}),
\end{equation}

\noindent
and also:

\begin{equation}
g({\rm J}){\rm B}_{ul}({\rm J}) = g({\rm J}-1){\rm B}_{lu}({\rm J}).
\end{equation} 

\noindent
The explicit equation for 
$\phi({\rm v}_z)$ is given by:

\begin{equation} 
\phi ({\rm v}_z) = \frac{c}{\sigma_{\rm v}}\nu_{\rm J} \sqrt{2\pi}\exp 
\left(-\frac{{\rm v}_z^2}{2\sigma_{\rm v}^2}\right),
\end{equation}

\noindent
and the number density $n({\rm J})$ of tracer molecules in the level J may
be evaluated by:

\begin{equation}
n({\rm J}) = n_{\rm tot}\frac{(2{\rm J}+1)
\exp (-h\nu_{\rm J}/kT_{\rm kin})}
{\sum_{{\rm J}^{\prime}=0}^{\infty}
(2{\rm J}^{\prime}+1)\exp (-h\nu_{{\rm J}^{\prime}}/kT_{\rm kin})}.
\end{equation}

\noindent
The sum in the denominator is the partition function, and
usually converges rapidly after about ten or so 
terms (depending on $T_{\rm kin}$), and can be easily evaluated to the 
required accuracy. 

Substituting these expressions into equation~9, 
with $I_{\nu}^{\rm ext}({\rm J})=B_{\nu}(\nu_{\rm J},2.73K)$, 
gives an expression for
the predicted intensity as a function of ${\rm v}_z$. 
To compare these predictions
with the Stenholm code, we used a static model cloud with a uniform kinetic 
temperature, hydrogen number density, velocity dispersion, 
and abundance. The hydrogen
number density was set several orders of magnitude above the critical density 
to enforce LTE. 

Figure~1 shows comparisons of the analytical and numerical 
predictions for the CS(J$=5\rightarrow 4$) line, from LTE models with 
progressively increasing CS abundance, covering a wide range of cloud 
optical depths.
The excellent agreement between the predicted spectra in all parts of 
the line, and
across the large range of optical depths, is a good indication that the
integration of the equation of transfer along the line of sight is being
carried out correctly. 

\subsection{Non-LTE comparison}

The LTE comparison shown above, while instructive, does not test the 
ability of the code to carry out non-LTE radiative transfer calculations 
reliably.
We have not found any exact analytical solutions for non-LTE radiative 
transfer 
problems in spherical geometry, to compare with the code in a similar 
manner as shown
above. Instead, we attempt to reproduce the line profiles obtained in 
a previous 
non-LTE radiative transfer study (Kr\"ugel \& Chini 1994, hereafter KC), 
which used
a somewhat different numerical method.

Figure~2 shows a comparison of profiles taken from KC with
the corresponding profiles calculated by our Stenholm code. 
The figure caption gives details of the model cloud used, 
which has strong velocity
and density gradients towards the centre, and is far from LTE. 
Both calculations use the CS collision rates 
of Green \& Chapman (1978).
Overall, the agreement between the two sets of line profiles is very good.
The intensities of the line profiles predicted by Stenholm 
are up to 10\% weaker
than the KC profiles, and the disagreement appears to be greatest 
for the smallest beam sizes. 

We checked our calculations, and found
that this was not the result of inadequate sampling of the lines of sight 
used in the beam convolution, by re-calculating the profiles 
for a range of different beam sampling densities. The small disagreement in
the line profile strengths is not serious, and may be caused by differences
in the spatial, angular or frequency discretisation schemes used in the 
models, or small differences in the precise values of some of the parameters 
used. Particularly encouraging is the very good agreement in the shapes 
of the corresponding line profiles.

\section{Approach to the modelling}

In this section we discuss our approach to the modelling of infall line
profiles, and investigate the qualitative dependence of the
predicted line profiles on a number of different model parameters.
The approach to the modelling is necessarily a compromise between 
the desire to represent realistically the object being modelled, and the
practical need to limit the parameter space and running time of the 
numerical simulations. The observations are modelled using the spherically 
symmetric micro-turbulent $\Lambda$-iteration radiative transfer code 
described above. Even with the considerable 
simplification of spherical symmetry the number of model
parameters is formidable. 

Radial profiles of 
density, kinetic temperature, molecule abundance, systematic velocity, 
and micro-turbulent velocity width must be specified. Additional 
parameters include 
the choice of inner and outer boundaries of the model, the incident radiation
fields at the inner and outer boundaries, and the distance to the object. 
Some assumptions must be made about the forms taken by 
these profiles, 
so that they may be described by specifying only a few parameters. 
Some of the profiles are assumed to have 
a power law dependence with radius, which requires only the power law
exponent and a normalisation factor to be specified. Wherever
possible, the choice of parametric forms used to describe the radial
profiles is guided by theoretical considerations. 
We now discuss the profile of each physical parameter in turn. 

\subsection{The systematic velocity profile}

Following the onset of fully dynamical collapse, the radial velocity profile 
should tend towards a freefall profile 
(e.g. Larson 1969; Shu 1977; Hunter 1977; Zhou 1992; 
Foster \& Chevalier 1993). The pure free-fall velocity profile may be 
written as:

\begin{equation}
v_r(r) = -\left(\frac{2GM}{r}\right)^{\frac{1}{2}}.
\end{equation}

\noindent
The infall velocity profiles in non-rotating clouds
which collapse from different initial conditions should approximately 
converge, even when the mass accretion rates are very different, as 
long as the comparison is made 
at times when the central protostellar masses are equal (Hunter 1977).
Hence we use the singular isothermal sphere (SIS) model velocity profiles to parameterise
the infall velocity field
in the radiative transfer modelling, which allows the velocity 
field to be described in terms of 
only two quantities: the effective sound speed $a_{\rm eff}$ and the
infall radius $r_{\rm inf}$. We stress that these are merely used as 
parameters characterising the velocity field,
and their actual physical significance is model dependent. 

We have found the following empirical fit to the exact velocity 
profile predicted by the 
standard SIS model, for $r<r_{\rm inf}$:

\begin{equation}
\frac{v_r(r)}{a_{\rm eff}} \simeq -\sqrt{2}\left(\frac{r}
{r_{\rm inf}}\right)^{-\frac{1}{2}} + 
\sqrt{2}\left(\frac{r}{r_{\rm inf}}\right)^{0.15}.
\end{equation}

\noindent
This equation fits the SIS collapse field to within a few per cent.
If the collapsing cloud has some initial rotation, then
centrifugal support may become important at small radii, leading to a
decrease in the radial velocity close to the centre of collapse.
With an initial angular velocity $\Omega$, then the inner radius, $r_d$, 
at which centrifugal forces begin to dominate depends on the mass 
already accreted, $M_{\rm acc}$, 
and the radius $R$ enclosing this mass (Terebey, Shu \& Cassen 1984)
through the following relation:

\begin{equation}
r_d = \frac{\Omega^2R^4}{GM_{\rm acc}}.
\end{equation}

\subsection{Micro-turbulent velocity profile}

As discussed above, 
the radiative transfer code used to model the data in this paper makes use
of the micro-turbulent approximation. This assumes that random
`turbulent' motions can be treated in the same way as thermal
molecular motions, by incorporating them into the local line 
profile function. 
Comparisons between radiative transfer codes which model 
macroscopic and microscopic turbulence suggest
that the main effect of relaxing the micro-turbulent approximation 
is to reduce the strength of self-absorption features 
(e.g. Park \& Hong 1995).

Although observations appear to indicate random motions of some kind 
must be present in protostellar envelopes, 
very little theoretical work has been done on the role of turbulence 
and turbulent support in protostellar
collapse. Lizano \& Shu (1989) derived a `turbulent equation of 
state' using 
the well known `Larson's Laws' (Larson 1981), which relate the observed 
velocity dispersion 
($\sigma$), size (R), and average density ($\bar{\rho}$) in molecular 
clouds and cores:

\begin{eqnarray}
\sigma & \propto & R^{\frac{1}{2}}, \label{lws}\\
\bar{\rho} & \propto & R^{-1}.
\end{eqnarray}

\noindent
These correlations apply to ensembles of clouds and also to 
observations of individual clouds 
in different molecular tracers, and are associated with the 
turbulent motions
in the clouds (e.g. Myers 1983). Combining the two relations above gives:

\begin{equation}
\sigma \propto \bar{\rho}^{-\frac{1}{2}}.\label{sigrho}
\end{equation}

\noindent
Zhou (1992) used this relation to infer the variation of turbulent
velocity with density in his radiative transfer calculations of
infall line profiles.

V\'azquez-Semadeni, Cant\'o \& Lizano (1998) have recently argued that 
Larson's Laws may be inappropriate to dynamically collapsing
clouds, because they were established from observations of clouds
which were in equilibrium configurations. These authors carried
out numerical hydrodynamic and magneto-hydrodynamic simulations to test
the behaviour of the turbulence during the collapse, and found that,
in contrast to equation~\ref{sigrho}, the turbulent velocity dispersion
actually increases with density during the collapse. 

For purely 
hydrodynamic and weakly magnetic collapse models the relation between
the turbulent velocity dispersion and the density was seen to 
approach a power law 
of the form $\sigma \propto \rho^{1/2}$ (i.e. $P_{\rm tb}\propto
\rho^2$). In the case of strongly magnetically inhibited
collapse, the power law index was observed to
change to $\sigma \propto \rho^{1/4}$ 
($P_{\rm tb}\propto \rho^{3/2}$), consistent with the predicted
behaviour for the slow compression of Alfv\'en waves (McKee \& Zweibel
1995). The behaviour of the turbulence in infalling protostellar envelopes 
is thus theoretically rather uncertain. Turbulence nevertheless
plays a very important part in the formation of line profiles, and is 
one of the principal sources of uncertainty in the
infall line profile modelling. 

\subsection{Density profile}

In a region where the infall velocity has a free-fall profile there is
often a strong tendency for the density profile to adopt an
$r^{-3/2}$ distribution, since if 
the rate of injection of material into the accretion stream is not changing 
rapidly with time, the quantity $4\pi r^2\rho v_r$ should be 
approximately independent
of radius. The prediction for the density profile deep inside the 
infalling region may be written as:

\begin{equation}
\rho (r) = \frac{a_{\rm eff}^2}{\pi\sqrt{32}\, 
Gr_{\rm inf}^{1/2}}\,r^{-\frac{3}{2}}, 
\end{equation}

\noindent
where, as before, $r_{\rm inf}=a_{\rm eff}t$ is the infall radius, 
and $a_{\rm eff}$ is the effective 
sound speed. The exact model density profile inside the whole
infalling region is very well approximated (within $\sim$1 per cent)
by the following empirical formula:

\begin{equation}
\rho (r) = \frac{a_{\rm eff}^2}{2\pi Gr_{\rm inf}^2}
\left[ 0.35\left(\frac{r}{r_{\rm inf}}\right)^{-\frac{3}{2}}
+ 0.65\left(\frac{r}{r_{\rm inf}}\right)^{-0.64} \right]. \label{shurho}
\end{equation}

\noindent
Outside the infall radius, the density profile is given by:

\begin{equation}
\rho (r)=\frac{a_{\rm eff}^2}{2\pi Gr^2}.
\end{equation}

\noindent
There are several competing theoretical models which predict 
different forms of the density profiles 
at various stages in the contraction and collapse of a dense 
cloud core (e.g. Foster \& Chevalier 1993; 
Basu \& Mouschovias 1994; Whitworth \& Ward-Thompson 2001). 

However, once a central protostar
has formed there is a tendency in the theoretical models for
the forms of the density and velocity 
profiles to converge towards the SIS model forms, 
in particular towards the centre of the collapse 
(e.g. Foster \& Chevalier 1993). We therefore 
initially search for fits to our observations using 
density profiles of the above form, 
although we do not require that the adopted 
density profile uses the same values for the parameters $r_{\inf}$ and
$a_{\rm eff}$ as used in the velocity profile.

\subsection{Kinetic temperature profile}

The temperature of the gas in an infalling protostellar envelope is 
determined by the balance
of heating and cooling processes. Ceccarelli, Hollenbach \& Tielens 
(1996) modelled these processes 
to derive the temperature profile in protostellar envelopes with
central embedded heating sources.
They found that in all the cases studied, the gas temperature 
profile stays within 
$\sim$30\% of the dust temperature, usually lying slightly below 
the dust temperature, $T_d$. In the following 
it is assumed that the gas temperature is well coupled
to the dust temperature, although 
this assumption may be significantly in error for 
hydrogen number densities below $\sim 10^4$\,cm$^{-3}$. 

If the emitted radiation reaching the optically thin part of the 
envelope peaks at a wavelength 
$\lambda_{\star}$, then the temperature of the dust at a radius $r$ will be given by:

\begin{equation}
T_d(r) \propto \left( \frac{L_{\star}}
{r^2\lambda_{\star}^{\beta}}\right)^{1/(4+\beta)}. \label{tlaw}
\end{equation}

\noindent
The value of $\lambda_{\star}$ depends on the details of the dust 
radiative transfer through the optically 
thick inner region (i.e. how the material in this region is distributed). 
Kenyon et al. (1993) used the diffusion approximation in the optically 
thick region to estimate the 
value of $\lambda_{\star}$. For a density profile of the form 
$\rho (r) = \rho_0 r^{-q}$ this gives: 

\begin{figure*}
\setlength{\unitlength}{1mm}
\begin{picture}(80,90)
\includegraphics{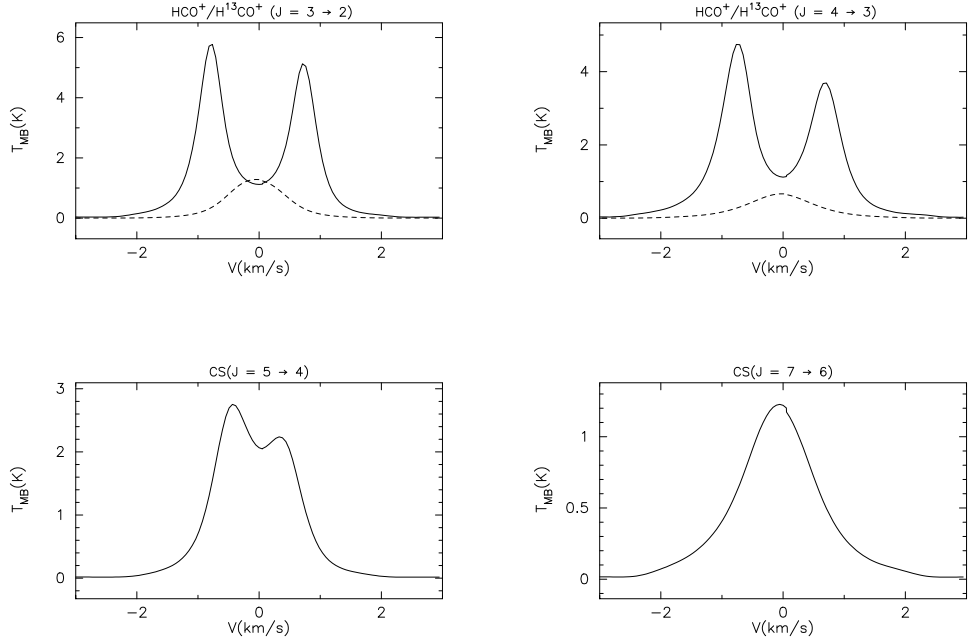}
\end{picture}
\caption{Predicted JCMT line 
profiles for the canonical infall model
(see text for details) for an assumed distance of 200\,pc. The
rare isotopomer lines are plotted as dashed lines in the top panels.} 
\label{can1}
\end{figure*}

\begin{equation}
\lambda_{\star} \propto \left[ \frac{\rho_0\, ^{2/(q-1)}}{L}
\right]^{1/(4+4\beta)}.
\end{equation}

\noindent
Substituting this expression into the above equation and
setting $\beta=1.5$ and $q=1.5$ (appropriate
for a hydrodynamically collapsing region), we find:

\begin{equation}
T_d(r) \propto \rho_0^{-0.055} L_{\star}^{0.21} r^{-0.36}. \label{tlaw2} 
\end{equation}

\noindent
The temperature profile in the optically thin part of the 
envelope is therefore only weakly 
dependent on the density normalisation and the luminosity
(see Kenyon et al. 1993 for further discussion).

For the objects modelled in this paper, the optically thick 
region is expected to be very much smaller than the beam size
of the observations and is comparable to the inner shell 
radius used in the radiative transfer models. 
At sufficiently large radii, the gas temperature will 
approach the ambient temperature of the larger scale cloud. 
We therefore adopt the temperature profile appropriate to 
the optically thin limit ($T\propto r^s$ where 
$-0.4 < s < -0.33$) in the inner region, 
and switch to a flat profile at the radius where the power law 
inner profile falls below the ambient cloud temperature. 

\begin{figure*}
\setlength{\unitlength}{1mm}
\begin{picture}(80,40)
\includegraphics{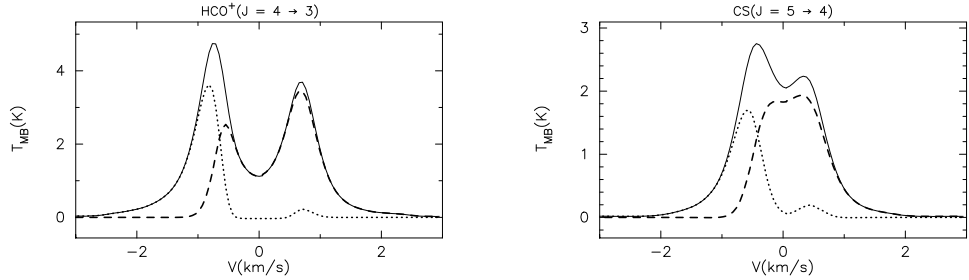}
\end{picture}
\caption{Plots showing the contribution of the emission from the
envelope hemisphere nearer to the observer
(dashed line) and the attenuated emission from the further half of the 
envelope (dotted line), to the predicted \hcoft and \csff line profiles 
(solid lines).} 
\label{modex1}
\end{figure*}

\begin{figure*}
\setlength{\unitlength}{1mm}
\begin{picture}(80,40)
\includegraphics{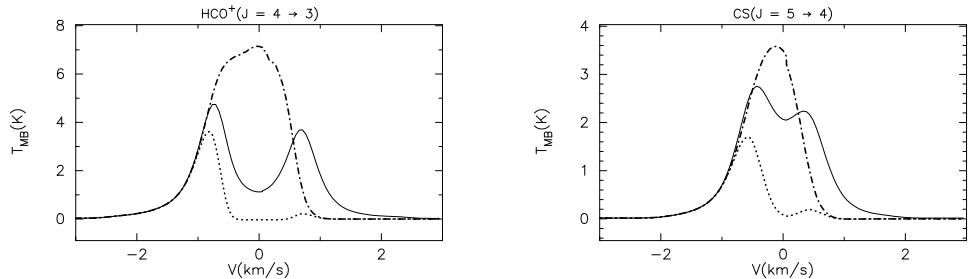}
\end{picture}
\caption{Plots showing the unattenuated emission from 
the further half (from the observer) 
of the model envelope (dot-dashed line) in the 
\hcoft and \csff lines. 
The full line profile (solid line) 
and the attenuated emission from the further half 
(dotted line) are shown for comparison.} 
\label{modex2}
\end{figure*}

\subsection{Abundance profile}

The profiles of CS and \hco relative abundance in an infalling
protostellar envelope are governed by time dependent chemistry 
and the transfer of 
molecules between the gas phase and the ice mantles surrounding dust grains. 
This problem has been studied by Rawlings et al. (1992) and Willacy, 
Rawlings \& Williams (1994) in the case of isothermal collapsing clouds. 
The objects studied in this paper contain internal stellar heating sources,
and are unlikely to be well described by an isothermal model. 
The chemistry of internally heated infalling protostellar envelopes has
been studied by Ceccarelli, Hollenbach \& Tielens (1996). We refer the reader to all of these references for further discussion of the way in which chemistry can affect observed abundance profiles.

Given these uncertainties  
and the requirement to limit the parameter space,
we used flat abundance profiles in our modelling. 
The fractional abundances of CS and HCO$^+$ measured in dense 
molecular gas cores occupy similar ranges, typically between 
$10^{-9}$ and $10^{-8}$ (e.g. Irvine et al 1987;
McMullin, Mundy \& Blake 1994; Blake et al. 1995). The measured values of 
[$^{12}$C/$^{13}$C] 
at the galacto-centric radius of the sun (which applies to all of 
our observed objects) 
lie in the range 50--100 (Wilson \& Rood 1994). We therefore restrict the 
abundances used in the
modelling to these ranges of values.

\begin{figure*}
\setlength{\unitlength}{1mm}
\begin{picture}(80,40)
\includegraphics{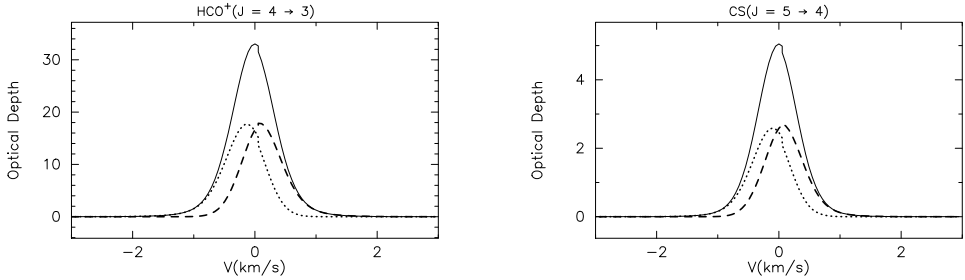}
\end{picture}
\caption{Plots showing the contribution to the total 
beam-averaged optical depth through the envelope (solid line) 
from the near and far hemispheres of the canonical 
infall model (dashed and dotted lines respectively), in the \hcoft and 
\csff lines.} 
\label{canopt}
\end{figure*}

\begin{figure*}
\setlength{\unitlength}{1mm}
\begin{picture}(80,40)
\includegraphics{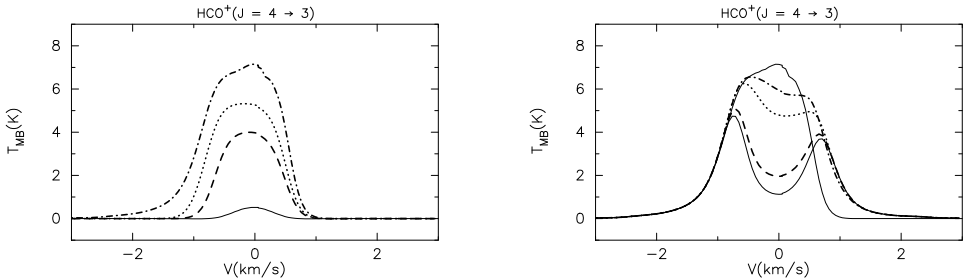}
\end{picture}
\caption{Plots illustrating how the \hcoft evolves as 
progressively more of the cloud is
included in the line profile calculation. The profiles 
represent different `slices' through 
the model cloud, where the slice planes are perpendicular 
to the line of sight and only the part 
of the cloud which lies beyond the slice plane is included. 
The left hand plot shows profiles obtained
for slice planes at various positions in the far half of 
the cloud. The distances
of the slice planes from the cloud centre along the line 
of sight are: 5000\,AU (solid line);
2000\,AU (dashed line); 1000\,AU (dotted line); and 0\,AU 
(dot-dashed line). The right hand plot shows the profiles
obtained for slice planes lying in the near half of the 
cloud. The distances of the planes from the cloud centre
are 0\,AU (solid line); 1000\,AU (dot-dashed line); 2000\,AU 
(dotted line); 5000\,AU (dashed line); and 10000\,AU 
(double-peaked solid line). The radius of the cloud is 10000\,AU.} 
\label{bbb}
\end{figure*}

\section{Parameter study of infall profiles}

\subsection{The canonical infall model}

It is useful to define a specific infalling envelope model, with 
plausible physical parameters,
which may be used to investigate some general properties of infall 
spectral line profiles. 
We use the above density and velocity profiles 
at a stage when a mass M$_{\odot}$/2 has already accreted onto the 
central protostar, and a further 
$\simeq$M$_{\odot}$/2 of envelope gas is infalling towards it. 
If we choose an effective sound speed 
of $a_{\rm eff}=0.35\,{\rm km\, s}^{-1}$, 
then an infall radius of 
$\simeq$3700\,AU is implied. We therefore use the following model 
relations for the 
systemic velocity and hydrogen number density profiles:

\[ v_r = 0.49 \left( \frac{r}{3700AU}\right)^{0.15} 
-0.49\left(\frac{r}{3700AU}\right)^{-0.5}{\rm km/s}, \]

\[ \frac{n_{H_2}}{2\times 10^5/cm^3} =
0.35 \left(\frac{r}{3700AU}\right)^{-\frac{3}{2}}
+0.65\left(\frac{r}{3700AU}\right)^{-0.64} \]

\noindent
and outside the infall radius the systematic velocity is set to zero.
The density profile is given by:

\begin{equation}
n_{{\rm H}_2}(r) = 2.0\times 10^{5}\left(\frac{r}
{3700\,{\rm AU}}\right)^{-2}.
\end{equation}

\noindent
We truncate the density profile at an outer radius of 10000\,AU, which 
encloses a total mass of 2.75\,M$_{\odot}$. This corresponds to the 
typical radius at which 
dense cores merge with the more diffuse gas in the ambient molecular 
cloud.

The adopted
micro-turbulent velocity dispersion is $\sigma_{\rm tb}=0.3$\,km\,s$^{-1}$, 
which is assumed to 
be spatially constant. Although there is little empirical or physical 
justification for a flat turbulent velocity profile
(see the above discussion), we make this assumption 
as a zeroth order approximation, in the
absence of any clear consensus as to the correct description of 
turbulence in infalling envelopes. 

As discussed above, the kinetic temperature profile 
(assumed to be equal to the dust temperature profile) 
in the optically thin part of the envelope is expected to be 
fairly insensitive to the 
luminosity of the central source. We choose a canonical 
temperature profile of:

\begin{equation}
T(r) = 20\,\left(\frac{r}{1000{\rm AU}}\right)^{-0.36}\ {\rm K}.\label{tcan}
\end{equation}

\noindent
The normalisation of the temperature profile was fixed using the dust 
temperature profile calculated
by Kenyon et al. (1993), for a source luminosity of 
$\sim 15 L_{\odot}$. 
The varying kinetic temperature in the envelope is not strictly 
consistent with the assumption of 
constant $a_{\rm eff}$ implied by the SIS model, where 
$a_{\rm eff}^2=\sigma_{\rm tb}^2+kT/\bar{m}$.
However, our primary concern is not to investigate the SIS model 
solution in detail, but rather to investigate the 
essential features of line formation in infalling envelopes, 
whether the infall is described 
by the SIS model or not. 

We set both the CS and \hco relative abundances to $5.0\times 10^{-9}$, 
and assume a 
[H$^{12}$CO$^{+}$]/[H$^{13}$CO$^+$] isotopic
ratio of 60. The set of molecular constants used in the radiative transfer 
modelling are listed in Table~1, where $\mu$ denotes
the permanent molecular dipole moment and other symbols take their usual
meanings. The data are taken from the JPL spectral 
line catalogue (Poynter \& Pickett 1985). 

\begin{table}
\label{molecules}
\begin{center}
\begin{tabular}{||c|c|c|c|c|} \hline
Molecule & Mass & $\mu$ & B$_{\rm rot}$ & hB$_{\rm rot}$/k \\ 
& (10$^{-27}$kg) & (10$^{-30}$C\,m) & (GHz) & (K) \\ \hline
HCO$^+$ & 48.16 & 12.2 & 44.59 & 2.14\\
H$^{13}$CO$^+$ & 49.81 & 13.0 & 43.38 & 2.08\\
CS & 73.02 & 6.54 & 24.50 & 3.50\\ \hline 
\end{tabular}
\end{center}
\caption{Assumed physical constants of the molecules modelled.}
\end{table}

\begin{figure*}
\setlength{\unitlength}{1mm}
\begin{picture}(80,40)
\includegraphics{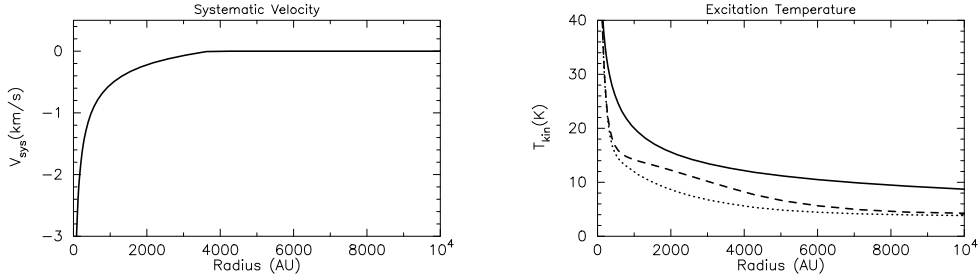}
\end{picture}
\caption{The left hand panel shows the radial profile of systematic velocity 
for the canonical infall model. In the right hand panel, the solution
for the excitation temperatures of the \hcoft (dashed line) and
\csff (dotted line) transitions are plotted. The kinetic temperature
profile (solid line) is also plotted for comparison.} 
\label{radplot}
\end{figure*}

\begin{figure*}
\setlength{\unitlength}{1mm}
\begin{picture}(80,40)
\includegraphics{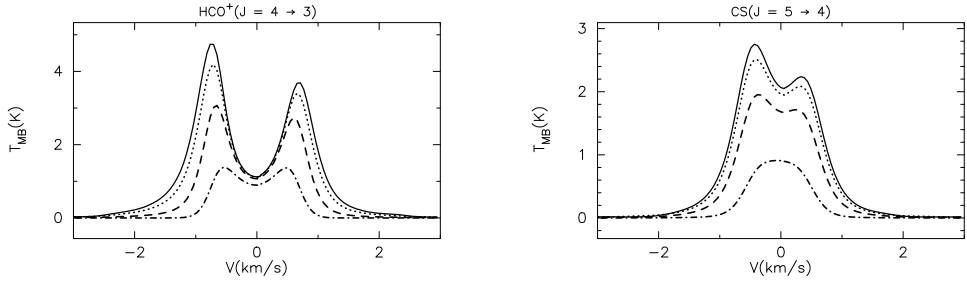}
\end{picture}
\caption{Plot showing the dependence of the line profiles
on the impact parameter of the beam direction relative to the centre of the
cloud. The solid lines show the 
on-source line profiles. The line profiles
for impact parameters of 1000\,AU (dotted line), 2000\,AU (dashed line), and 
4000\,AU (dot-dashed line) are also plotted. The assumed distance to the 
cloud is 200\,pc.}
\label{imppar}
\end{figure*}

\subsection{The origin of the infall profile}

We now use the radiative transfer code described above to simulate
observed spectral line observations of the canonical infall model, and 
investigate how the predicted line profiles depend on a number of 
different model parameters.
Figure~3 shows the predicted \hco , \htco and CS line profiles for the 
canonical infall model described above. The model cloud was assumed to 
lie at a distance of 200\,pc, and the 
appropriate beam sizes for each transition
were used when carrying out the beam convolution.

Double-peaked, blue-asymmetric line profiles are seen in all of the 
main isotopomer transitions, 
apart from the \csss line, which is nevertheless skewed slightly bluewards
of the systemic velocity. The minima between the peaks of the double-peaked 
profiles lie
close to the systemic velocity. The \hcott line shows a stronger 
self-reversal 
than the \hcoft line, and in both of these lines the self-reversal is 
stronger than the \csff line. 
The two single-peaked \htco lines peak very close to the systemic velocity,
coinciding with the minima in the main line profiles. Low level high-velocity 
wings are visible in all of the main line profiles, and appear to be 
stronger in the
higher rotational transitions of both molecules.

To illustrate how the asymmetries in line profiles arise, we show in 
Figure~4 the separate contributions of the near and far 
portions of the envelope to the line profiles for {\mbox the \hcoft} 
and \csff transitions. 
The nearer hemisphere (to the observer)
of the envelope (in which the systematic velocity is 
red-shifted) is seen to be mostly responsible for the red-shifted peak 
in the line profile, as expected. However, the
more surprising observation is
that the nearer hemisphere also dominates the emission at velocities 
well into the blue-shifted 
side of the line profile, and contributes significantly to the 
blue-shifted peak. 

Emission from the far
hemisphere only becomes dominant in the most blue-shifted part of the line. 
It is interesting to note that the contribution of the far hemisphere
to the blue peak is no greater than the contribution of the near
hemisphere to the red peak, and 
it is the extra contribution of the near hemisphere to the 
blue-shifted emission which produces the 
blue asymmetry. This contrasts with the explanation usually given 
for the infall asymmetry, in which it is argued
that the optical depth to the warm dense blue-shifted gas just 
beyond the central protostar is less than the optical 
depth to the red-shifted gas just in front of the protostar, 
due to the Doppler shift of the intervening infalling envelope.
If this explanation were the correct one, then we would expect 
the blue-shifted emission from the far hemisphere to be, 
by itself, much brighter than the red-shifted emission from 
the near hemisphere, which contradicts our findings.

Figure~5 shows how the 
emission from the far half of the envelope is modified 
by the optical depth of the near half of the envelope 
on its way to the observer. Self-absorption clearly plays 
a very important role in shaping the emergent line profile.
In Figure~6 the contribution of the beam-averaged 
optical depths through both hemispheres of the envelope for both 
\csff and \hcoft are 
plotted. The total optical depth of the \hcoft line is much 
greater than
the \csff line, which accounts for the stronger self-absorption 
features in the 
\hco line. As expected, there is a separation in velocity
of the peaks of the optical depth profiles from the near and far 
hemispheres, 
although this is rather small ($\sim 0.2$\kms).

\begin{figure*}
\setlength{\unitlength}{1mm}
\begin{picture}(80,40)
\includegraphics{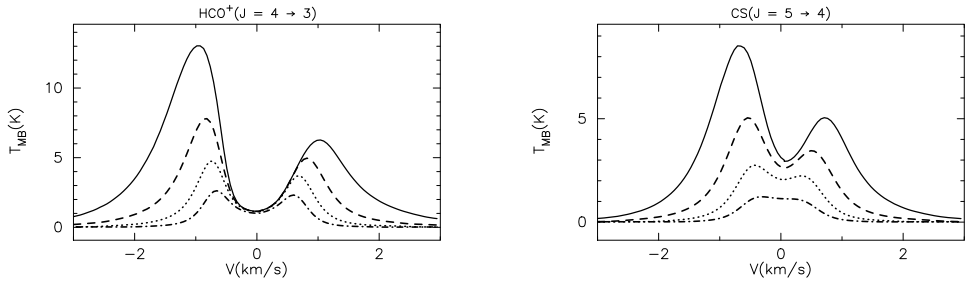}
\end{picture}
\caption{Plots showing the dependence of the predicted line profiles on the
assumed distance to the cloud. Spectra are plotted for distances of 
50\,pc (solid lines), 100\,pc (dashed lines), 200\,pc (dotted lines), 
and 400\,pc (dot-dashed lines).}
\label{dvar}
\end{figure*}

\begin{figure*}
\setlength{\unitlength}{1mm}
\begin{picture}(80,40)
\includegraphics{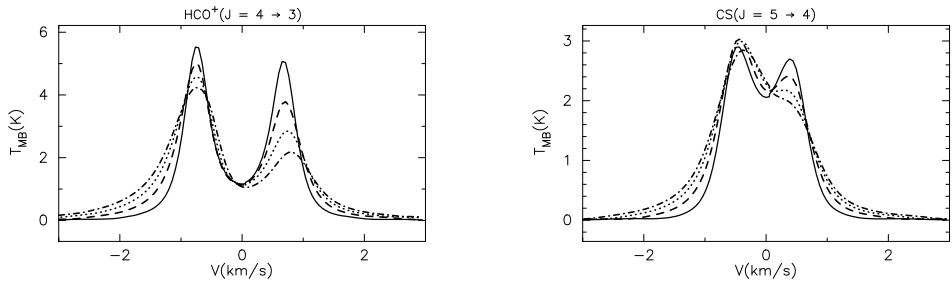}
\end{picture}
\caption{Plots illustrating the dependence of the line 
profiles on infall velocity. The infall velocity profiles were
calculated for infall radii of 2000\,AU (solid lines),
4000\,AU (dashed lines), 6000\,AU (dotted lines) and 8000\,AU 
(dot-dashed lines). 
The canonical values were used for all other model parameters.}
\label{canv}
\end{figure*}

To shed further light on how the asymmetric infall profiles are 
formed, we show in Figure~7
how the \hcoft line profile changes as successive `layers' of 
the model envelope are included, from
the far edge of the envelope forwards. For all the `slice planes' 
in the rear of the envelope (left hand
panel of Figure~7) the predicted profiles are single peaked. 
This is explained by the fact that in the far half of the envelope,
the higher excitation gas always lies in front of the lower excitation 
temperature gas. 
Figure~8
shows the radial dependence of the systematic velocity, kinetic 
temperature, and 
\hcoft and \csff excitation temperature in the model. The fact 
that the excitation 
temperature lies well below the kinetic temperature over most 
of the envelope 
is a demonstration of the inapplicability of the LTE 
approximation to this problem.

As successive layers of the front half of the envelope are 
included in the line profile calculation 
(right hand panel of Figure~7), there is a reduction in the intensity in
the core of the line. This is due to the fact that in the near 
half of the cloud, lower excitation temperature gas
lies in the foreground, and absorbs the emission from the higher 
excitation gas behind it. 
Nevertheless, the emission of this gas is also very important in 
determining the form of the
emergent line profile. The emission from gas near the centre,
in the near half of the envelope, is seen in Figure~7
to produce an increase in the 
line intensity at strongly red-shifted velocities. The contribution 
of this gas to the emission at
lower velocities is obscured in the figure by the dominating effect 
of absorption 
towards the line centre, but Figure~7 shows that this 
emission is significant 
even on the blue-shifted side of the line. In the outermost region 
of the envelope,
where the excitation temperature is lowest, the gas is predominantly 
absorbing,
and simply deepens the self-reversal of the line profile.

\subsection{Impact parameter and beam size}

Figure~9 illustrates how the predicted \hcoft and \csff line 
profiles depend on the impact parameter of the line of sight. 
As expected, the strength of the lines diminishes with increasing
impact parameter, as the beam samples increasingly less dense and
lower temperature gas. The degree of asymmetry in
the line profiles decreases with increasing impact parameter, 
partly because of the lower infall velocities at larger radii, and
partly because of the lower column densities, and hence optical depth, 
along the line of sight. The lines also 
become narrower with increasing impact
parameter, as a result of the decreasing infall velocity with radius.

The distances to many nearby protostellar objects are often only 
known to an accuracy 
of $\sim$50\%, so it is important to examine 
how sensitively the predicted line profiles depend on the assumed distance. 
There is of course a degeneracy between the assumed distance to the cloud 
and the beam size.

\begin{figure*}
\setlength{\unitlength}{1mm}
\begin{picture}(80,40)
\includegraphics{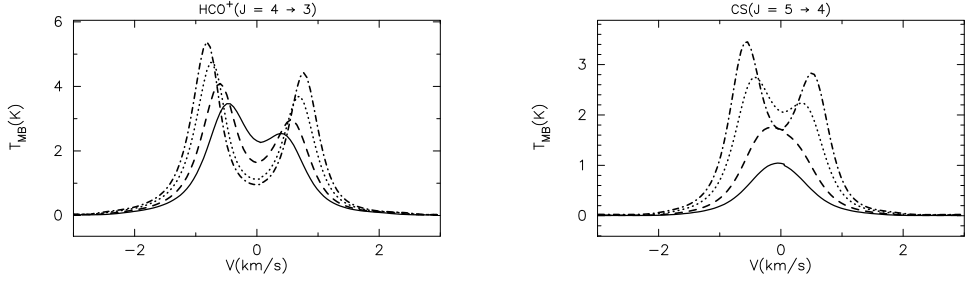}
\end{picture}
\caption{Plots showing the dependence of the predicted line profile on the
relative abundance of the tracer molecule. Spectra are plotted for 
HCO$^+$ and CS abundances (relative to molecular hydrogen) of $10^{-9}$ 
(solid lines), 
$2\times 10^{-9}$ (dashed lines), $5\times 10^{-9}$ (dotted lines), 
and $10^{-8}$
(dot-dashed lines).}
\label{abund}
\end{figure*}
\begin{figure*}
\setlength{\unitlength}{1mm}
\begin{picture}(80,40)
\includegraphics{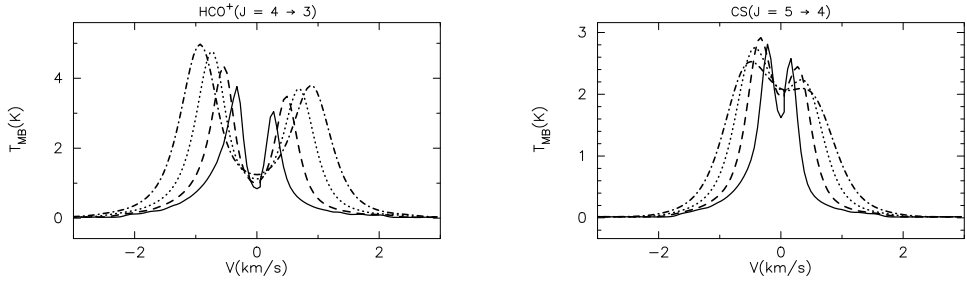}
\end{picture}
\caption{Plots showing the dependence of the predicted line profiles on the
magnitude of the turbulent velocity dispersion, assumed in this case to be 
constant throughout the 
envelope. Spectra are plotted for turbulent velocity dispersions of 0.1\kms
(solid lines), 0.2\kms (dashed lines), 0.3\kms (dotted lines) and 0.4\kms 
(dot-dashed lines). The equivalent
FWHM turbulent velocity widths are 
found by multiplying these numbers by 2.35.}
\label{turb0}
\end{figure*}

In Figure~10 we plot \hcoft and \csff line profiles for a number 
of different assumed 
distances, whilst keeping the beam sizes fixed.
The peak line temperatures increase approximately as the inverse 
of the assumed distance (or beam size). The increase in the strength of the
line wings depends approximately on the inverse square
of the assumed distance, which suggests that the 
line wing flux originates from 
a region on the source much smaller than the beam 
size. This is the explained by
the fact that in the velocity profile we have adopted, 
the highest infall velocities 
lie at the smallest radii, and therefore the high velocity 
wings of the line profile
are formed in a very small region around the centre of the 
cloud. The different 
dependence of the line core and line wings on distance produces 
increasingly broad profiles with decreasing distance and/or beam size.

We conclude that distance uncertainties may contribute significantly 
to the overall
uncertainty in the radiative transfer modelling of protostellar envelopes. 
Gregersen et 
al. (1997) analysed \hco observations of a large sample of Class 0 sources 
using a 
model cloud at a single `representative distance'. This
approach will not give reliable results when the scatter in the distances
of the objects in the sample is large.

\subsection{Infall velocity}

Figure~11 shows line profiles for a number of different 
infall radii, in which only the velocity profile
is varied, as discussed in section 3.1, and the density 
profile remains fixed as for the canonical model.
For increasing values of the infall radius, larger 
fractions of the envelope take part in the infall motion, and the collapse 
speed at a given radius increases approximately as $r_{\rm inf}^{1/2}$. 
The line profiles become more strongly skewed towards blue velocities 
with increasing infall velocity, mainly 
as a result of the diminishing intensity of the red-shifted peak. 
The actual velocities of the two peaks show
remarkably little variation over the range of infall velocities 
considered. Predictably, the models with the 
highest infall velocities produce the strongest emission 
in the line wings. In actual observations this infall 
signature may often be obscured by emission from the outflow. 

\subsection{Tracer molecule abundance}

Figure~12 shows how the predicted line profiles vary 
with the assumed relative 
abundance of CS and \hco in the envelope. The relative 
abundance of the tracer molecule
largely determines the optical depth through the envelope 
in the observed 
line, and is therefore a very important parameter in deciding 
the overall appearance of 
the line profile. This is apparent from the figure, where the 
intensities and shapes of the 
line profiles are seen to vary significantly over the factor of 
10 range of relative abundances 
covered. As expected, the lowest values of the relative abundance 
(and hence optical depth)
produce the weakest infall profiles. The progression of line 
shapes with increasing optical
depth is from: a single-peaked gaussian; to a single-peaked blue-skewed 
profile; to a profile
with a blue-shifted peak and red-shifted `shoulder' (or `red knee'); 
to a double-peaked
profile with a stronger blue-shifted peak. As the abundance and optical 
depth increase further,
the absorption trough deepens, the line peaks become stronger and their 
separation increases. 

\begin{figure*}
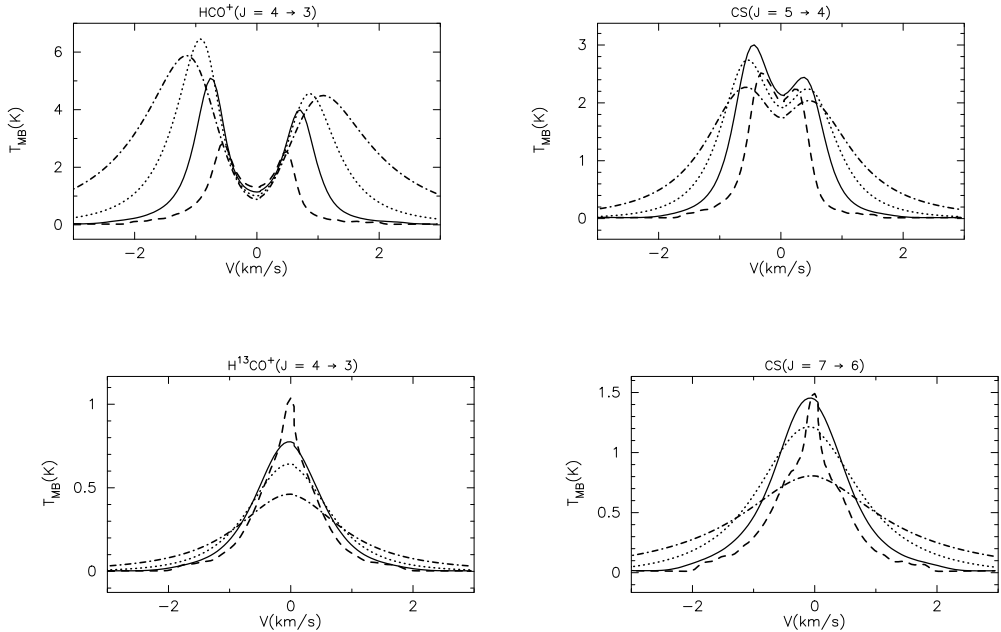

\setlength{\unitlength}{1mm}
\begin{picture}(80,90)
\includegraphics{fig14a.eps}
\includegraphics{fig14b.eps}
\end{picture}
\caption{Plots showing the predicted line profiles 
for different forms of the 
micro-turbulent velocity profile. The adopted relations 
between the turbulent 
velocity dispersion, $\sigma_{\rm tb}$, and the gas density, 
$\rho$, that were used are:
$\sigma_{\rm tb}= $constant (solid line); 
$\sigma_{\rm tb}\propto \rho^{-1/2}$ (dashed
line); $\sigma_{\rm tb}\propto \rho^{1/4}$ (dotted line); 
and $\sigma_{\rm tb}\propto \rho^{1/2}$ (dot-dashed line). 
Each profile was normalised such that the FWHM turbulent 
velocity width ($=2.35\sigma_{\rm tb}$) at the half-radius
of the cloud (5000\,AU) has a value of 0.3\kms.}
\label{turbb}
\end{figure*}

\subsection{Turbulence}

Figure~13 shows how the line profiles 
depend on the magnitude of the turbulent velocity
dispersion, $\sigma_{\rm tb}$, when it is assumed 
to be uniform throughout the envelope. 
As the turbulent velocity dispersion increases, 
the most apparent effect on the line profiles
is to increase the velocity separation between the 
two peaks, as a result of the broadening 
of the absorption profile of the foreground envelope. 
The total integrated flux in the 
line tends to increase with the turbulent velocity dispersion. 
This is because the peak optical depth of the 
foreground gas is reduced as the molecules are
spread over a wider range of velocities, and the similar 
reduction in the optical depth
of the strongly emitting gas in the centre of the cloud allows 
more of the emission
to `escape'. 

Comparing the \hcoft with the \csff profiles,
we see that the effect of changing the value of 
the turbulent velocity dispersion
depends strongly on the peak optical 
depth in the transition. When the 
optical depth is small, increasing the turbulence 
tends to `smear out' the double-peaked
profile, whereas for very optically thick lines the 
double-peaked structure remains 
just as promiment as the velocity gap between the two peaks increases.

Note that this finding is in apparent contradiction with some previous
models. For example, Myers et al. (1996) used a simple `two--slab' 
geometry to model infalling protostars, and claimed that as a protostar
evolves, the fraction of the envelope that is infalling increases (under
the SIS inside-out collapse assumptions), and that this causes
the gap between the two peaks to decrease, and the absorption trough to
become narrower. Our modelling indicates that {\em the amount of turbulence
is what actually determines the gap between the peaks}. The two scenarios
could only be reconciled if turbulence were to decrease as a protostar
evolves. This may seem unlikely, since the associated outflow is more likely 
to increase the turbulence as it evolves. However, a recent claim has been
made (Jayawardhana, Hartmann and Calvet 2001 -- hereafter JHC) 
that in fact Class 0 sources
may be formed preferentially in regions of higher turbulence than Class I
sources. This is an interesting idea that is consistent with our modelling,
although the other ideas of JHC that a significant fraction of a Class 0
envelope is not infalling, but rather static, is not consistent with our
modelling of specific Class 0 sources (see below).

Figure~14 illustrates how the line profiles vary with the 
exponent in the assumed
power-law relation between the turbulent velocity and the density. 
Guided by the discussion on turbulence in the section 3.2, 
we investigate turbulent 
velocity laws of the form $\sigma_{\rm tb}\propto \rho^{1/4}$ 
and $\sigma_{\rm tb}\propto \rho^{1/2}$ 
(as found in the numerical simulations of V\'azquez-Semadeni et al. 1998), 
and $\sigma_{\rm tb}\propto \rho^{-1/2}$ (derived using Larson's Laws). For
comparison we also plot the profile produced 
by the canonical model, which assumes
a flat turbulent velocity profile.

\begin{figure*}
\setlength{\unitlength}{1mm}
\begin{picture}(80,210)
\includegraphics{fig15.eps}
\end{picture}
\caption{Plot showing the effect of solid-body rotation 
on infall line profiles.
\hcoft and \csff line profiles are plotted in the left 
and right hand columns respectively. 
The solid lines show the predicted line profiles for the 
canonical infall model with
a projected angular velocity of solid-body rotation of
$\Omega \sin i$=30\kms \,pc$^{-1}$, where $i$ is the
inclination angle of the rotation axis to the line of sight. 
The lines are plotted for several 
impact parameters. Displacements from the centre of 
the cloud along a line perpendicular to the rotation 
axis in the plane of the sky are given in 
the title of each plot. Impact parameters with positive displacements
lie on the red-shifted side of the rotational velocity gradient. 
The dashed lines show the predicted line
profiles in the absence of rotation for comparison.}
\label{rotate}
\end{figure*}

\begin{figure*}
\setlength{\unitlength}{1mm}
\begin{picture}(80,50)
\includegraphics{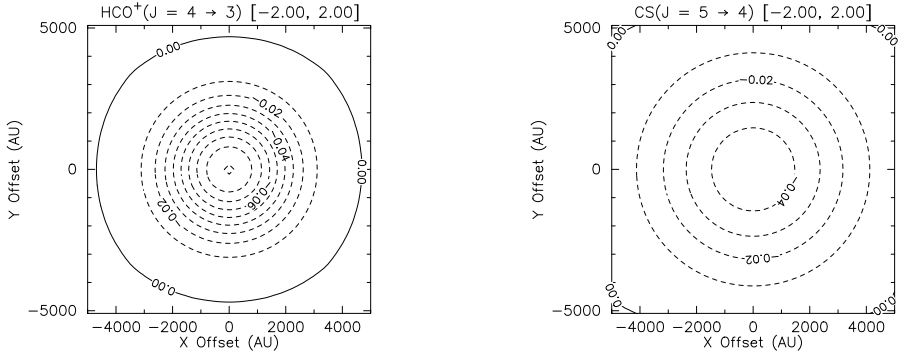}
\end{picture}
\caption{Simulated \hcoft and \csff centroid velocity 
contour plots (in \kms) produced from the 
canonical infall model with no rotation. 
The centroid velocity was calculated
over the velocity range $-$2.0 to $+$2.0 \kms.}
\label{cancvnr}
\end{figure*}

\begin{figure*}
\setlength{\unitlength}{1mm}
\begin{picture}(80,50)
\includegraphics{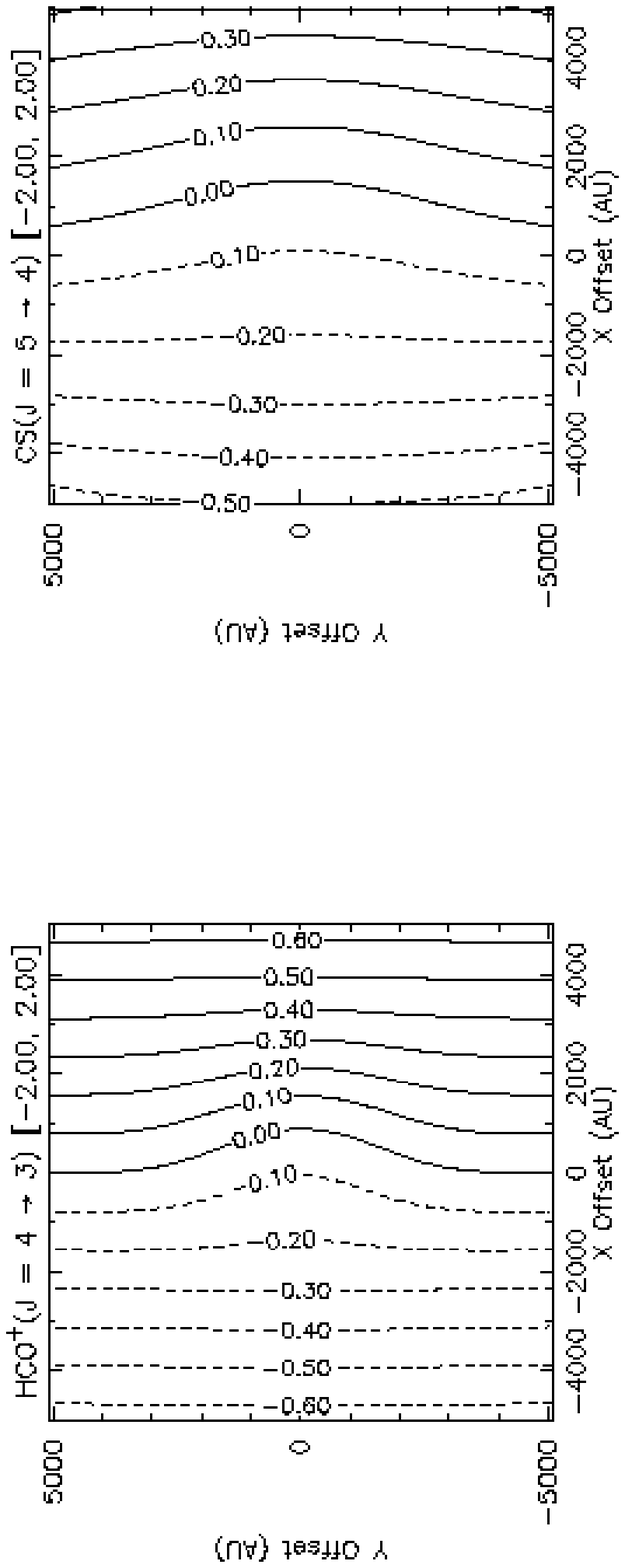}
\end{picture}
\caption{Simulated \hcoft and \csff centroid velocity 
contour plots (in \kms ) produced from the 
canonical infall model with a solid body rotation of 
$\Omega \sin i$=30\kms \,pc$^{-1}$, 
calculated over the velocity range $-$2.0 to $+$2.0 \kms.}
\label{cancv}
\end{figure*}

The profiles are clearly quite sensitive to the form of the 
turbulent velocity
law. The profiles obtained using the $\sigma_{\rm tb}\propto \rho^{-1/2}$ 
law (dashed
lines) are appreciably narrower than the other profiles. This
is particularly evident in the centres of the \htcoft and \csss 
lines. This
arises because the $\rho^{-1/2}$ law assigns the lowest turbulent velocity 
dispersion to
the hot, dense, inner regions of the cloud, where most of the line emission
takes place. These profiles also tend to show the strongest emission 
close to the line centre, mainly arising from the lower optical depth of the
outer envelope due to the enhanced velocity dispersion there. 
However, in the 
case of the \htcoft and possibly the \csss lines, this is due 
to the increased optical depth in the centre of the cloud, 
as these lines are not 
completely optically thick.

As the exponent $s$ in the relation $\sigma_{\rm tb}\propto \rho^s$ 
increases from
$-$0.5 to $+$0.5, the line profiles become much more dominated 
by broad line
wings. For the positive exponents, the largest velocity dispersions 
are located
in the centre of the cloud, which causes the emission from this 
region to be distributed
over a wide velocity range. The low turbulent and systematic 
velocities in the 
outer envelope confine the absorption of the gas in this region to line
centre velocities, allowing the enhanced high velocity emission from 
the core to remain 
virtually unobscured. Despite the change in overall appearance, the
\hcoft and \csff profiles retain the characteristic blue-asymmetric, 
double-peaked 
structure. The strong wings of these lines are similar to those normally 
associated
with outflow emission. The more optically thin \htcoft and \csss 
lines may provide
some means of distinguishing between these two possibilities. If the 
wings in the main line profiles are caused by enhanced turbulence 
near the centre of the cloud,
then the optically thin lines sould also show very broad profiles. 
Conversely, if the wings are tracing
a bipolar outflow, then the optically thin line profiles, which 
are expected to be dominated 
by the envelope, should be narrower.

\subsection{Solid-body rotation}

Figure~15 shows how the predicted 
line profiles change when a solid-body rotation is 
superposed on the infall velocity field.
This velocity field may not be physically
realistic, since angular momentum conservation will
tend to cause the angular velocity
of the gas to increase as it falls inwards, even in the 
presence of magnetic braking (e.g. Mouschovias 1994).
We use this example simply to illustrate some of the 
qualitative effects of rotation on line profiles.
A proper treatment of differential rotation would 
require a three-dimensional radiative transfer analysis, which is 
computationally a very large step from the 
spherically symmetric problem presented here.

The on-source line profiles are least affected by rotation, 
only showing a small decrease in the height of the peaks, 
and a very slight broadening of the line as a whole. For differential 
Keplerian ($\Omega \sin i \sim r^{-3/2}$) rotation, 
the on-source profile might
be more strongly affected, since the 
largest rotational velocities would then lie closest to the centre.

\begin{figure*}
\setlength{\unitlength}{1mm}
\begin{picture}(80,80)
\includegraphics{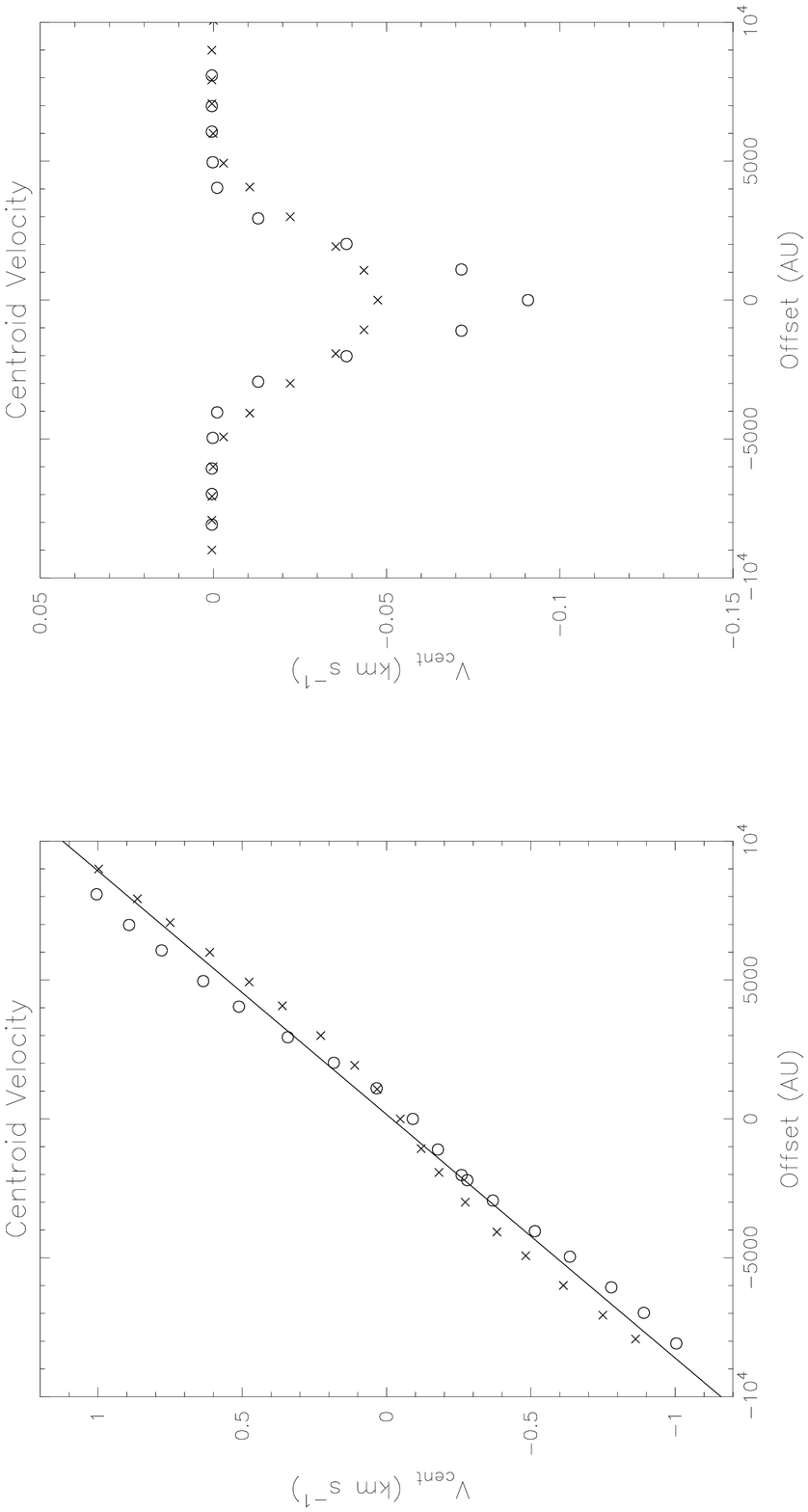}
\end{picture}
\caption{Centroid velocity measurements of the predicted 
\hcoft (open circles) and \csff (crosses) 
line profiles for the canonical infall model with a solid 
body rotation of $\Omega \sin i$=30\kms \,pc$^{-1}$.
The left hand panel shows a least-squares straight line fit 
to the centroid velocities measured along
a line perpendicular to the rotation axis (i.e. in the 
direction of maximum rotational velocity gradient)
passing through the centre of the cloud, while
the right panel shows the
centroid velocities along the rotation axis.} 
\label{cvplot}
\end{figure*}

The off-centre line profiles are more affected by solid-body rotation 
than the on-source profiles. 
As expected, the shifts in the centroid velocity of the line 
profiles follow
the rotational velocity gradient. As well as shifting the 
centroid, the rotation also significantly
distorts the shape of the off-centre line profiles. On the 
blue-shifted side of the rotational
velocity gradient, the rotation produces an enhanced blue 
asymmetry in the line profiles, 
in addition to the blue-asymmetry produced by the infall. 
At the positions on the red-shifted side of the 
rotational velocity gradient, the blue infall asymmetry is completely
reversed by the rotation.

Figures~16 \& 17 show \hcoft and \csff centroid velocity contour plots 
from the canonical infall model for projected rotational angular 
velocities of $\Omega \sin i$=0 and 30\kms \,pc$^{-1}$ respectively. 
The non-rotating, infalling model produces circularly
symmetric negative centroid velocity contours, which reach a 
minimum at the central position. 
The solid body rotation tends to produce approximately parallel 
contours of centroid velocity, 
aligned with the axis of rotation. 

However, the effect of the infall 
is to cause the contours to distort towards 
bluer velocities close to the centre of the cloud, 
producing the appearance a `blue bulge' in the centroid 
velocity contour plot, which encroaches into the red-shifted half of the 
rotational velocity 
gradient (Walker, Narayanan \& Boss 1994; Zhou 1995)

\begin{figure}
\setlength{\unitlength}{1mm}
\begin{picture}(60,100)
\includegraphics{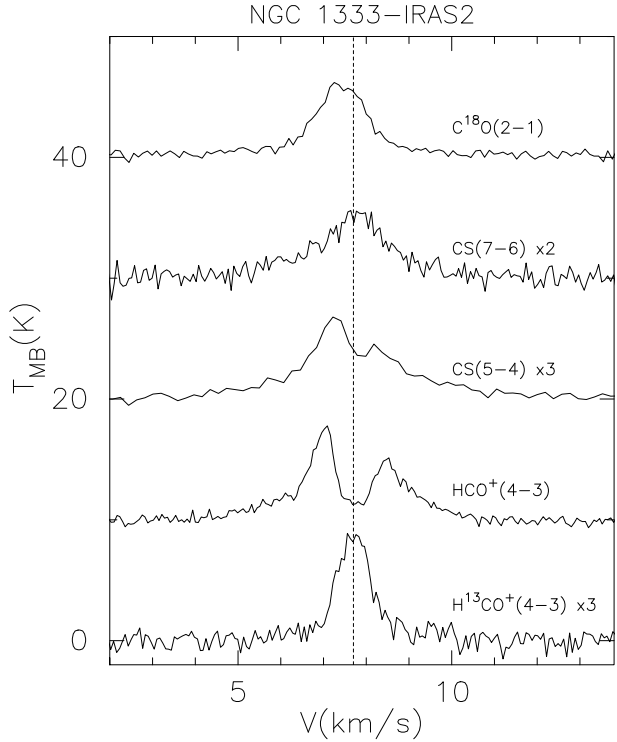}
\end{picture}
\caption{Spectra observed towards NGC1333--IRAS2. 
The vertical dashed line is at a velocity of 7.7\kms.
An extra 10K baseline shift has been added to each successive
spectrum.} 
\label{iras2spec0}
\end{figure} 

\begin{table}
\label{observe}
\begin{center}
\begin{tabular}{|l|c||} \hline
Date & Reference \\ \hline
1995 February 9--10 & Feb95\\
1995 July 26--28 & Jul95 \\
1996 March 17--22 & Mar96\\
1997 January 30 & Jan97\\
\hline
\end{tabular} 
\end{center}
\caption{JCMT observations and their references.}
\end{table}

We can use measurements of the 
centroid velocity gradient across our spectral line maps of
protostellar envelopes to place limits on the rotation rate 
of these envelopes. In Figure~18
the calculated \hcoft and \csff centroid 
velocities are plotted, along axes perpendicular to and parallel 
to the rotation axis, at 1000\,AU intervals (i.e. 5 arcsec
intervals for a distance of 200\,pc). 
A straight line fit to the centroid velocity along the perpendicular 
axis is shown in the left
hand panel of the figure. It is apparent that the CS centroid 
velocities lie behind
the overall velocity gradient on both sides of this gradient.

This is due to the larger assumed FWHM beam size
in the calculation of the \csff profile 
(19.3$^{\prime\prime}$ compared to 13.6$^{\prime\prime}$), 
which tends to increase the contribution of the brighter 
inner regions (which have lower rotational velocities) 
to the line profile. The centroid velocity along the 
axis parallel to the
rotation axis in the right hand panel clearly shows the `blue bulge' 
signature discussed above, although in this case the magnitude 
of the associated centroid velocity shift is rather small.

\begin{table}
\label{sources}
\begin{center}
{\footnotesize
\begin{tabular}{||l|c|c|c||} \hline
Source & R.A.(1950) & Dec.(1950) & D \\
Name & h\ \ \ m\ \ \ s & $^{\circ}\ \ \ ^{\prime}\ \ \ ^{\prime\prime}$ 
& (pc) \\ \hline 
N1333--IRAS2 & 03 25 49.9 & +31 04 16 & 220 \\
Serpens SMM1 & 18 27 17.3 & +01 13 23 & 300 \\
Serpens SMM4 & 18 27 24.7 & +01 11 10 & 300 \\
Serpens SMM3 & 18 27 27.3 & +01 11 55 & 300 \\
Serpens SMM2 & 18 27 28.0 & +01 10 45 & 300 \\
\hline
\end{tabular} 
}
\end{center}
\caption{Co-ordinates of the objects observed and their assumed distances.}
\end{table}

\begin{table*}
\label{frequencies}
\begin{center}
\begin{tabular}{||l|c|c|c|c|c|c||} \hline
Molecule & Transition & Frequency & FWHM & 
$\delta$v & $T_{\rm sys}$ & Reference \\
& & (GHz) & ($^{\prime\prime}$) & 
(km\,s$^{-1}$) & (K) & \\ \hline
HCO$^+$ & $J=3\rightarrow 2$ & 267.557625 & 18.3 & 
$\sim$0.25 & $\sim$ 500 & archive\\
& $J=4\rightarrow 3$ & 356.734248 & 13.6 & 0.080 & 
$\sim$ 800 & Jul95/Mar96\\
H$^{13}$CO$^+$ & $J=3\rightarrow 2$ & 260.25548 & 18.8 & 
$\sim$0.25 & $\sim$ 500 & archive\\
& $J=4\rightarrow 3$ & 346.998540 & 14.1 & 0.082 & 
$\sim$ 800 & Jul95/Mar96\\
CS & $J=5\rightarrow 4$ & 244.935606 & 19.3 & $\sim$ 
0.25 & $\sim$ 500 & Feb95/Jul95/Jan97 \\
& $J=7\rightarrow 6$ & 342.88294 & 14.2 & 0.083 
& $\sim$ 800 & Feb95/Jul95/Mar96\\
CO & $J=3\rightarrow 2$ & 345.795979 & 14.1 & 0.082 
& $\sim$ 800 & Mar96 \\
C$^{18}$O & $J=2\rightarrow 1$ & 219.56032 & 22.1 & $\sim$ 
0.25 & $\sim$ 500 & archive \\
C$^{18}$O & $J=3\rightarrow 2$ & 329.33050 & 14.8 & 0.087 & 
$\sim$ 1500 & Mar96 \\ \hline
\end{tabular} 
\end{center}
\caption{Table giving details of the submillimetre line transitions observed
using the JCMT. The line frequencies were taken from the Lovas spectral line
catalogue (Lovas 1992). The observing runs during which each transition was
observed are indicated in the last column (c.f. Table~2). Also
indicated are transitions which were obtained from the JCMT data archive. The
CO$(J=3\rightarrow 2)$ and CS($J=7\rightarrow 6$) transitions were observed
simultaneously, by selecting a frontend frequency which placed one transition
in each sideband.}
\end{table*}

The formal least-squares straight line fit to the 
centroid velocity gradient perpendicular to the rotation axis is 
$23.5\pm 6.8$\kms \,pc$^{-1}$, and
the intercept is $-$0.018\kms . The actual angular velocity
used in the model was $\Omega \sin i =30$\kms \,pc$^{-1}$, 
so the actual and measured values agree to within $1\sigma$.
The slightly low measurement for the centroid velocity gradient 
is mainly due to the lagging behind of the CS
centroid velocities mentioned above,
as can be seen from the figure. Nevertheless, this example shows that the 
centroid velocity gradient gives a reasonable estimate of the actual 
projected velocity gradient across
a cloud, at least for the case of solid body rotation.

\section{Observations}

The observations were carried out at the James Clerk Maxwell Telescope
(JCMT), using receivers RxB3i (Cunningham et al. 1992)
and RxA2 (Davies et al. 1992), with the Digital Autocorrelation
Spectrometer (DAS) backend.
Details of the observing runs are given in Tables 2--4. 
Pointing and focus checks were performed roughly once per hour
during each run. The focus was found to be very stable throughout, and the
pointing accuracy was $\sim 2$ arcsec. Observations of standard sources were
taken several times per night for each transition observed. We estimate the
absolute calibration is accurate to within $\sim$25\%, although the relative
calibration between different observations is much better than this.
Throughout the observations the DAS was configured with the optimum frequency 
resolution of 95kHz per channel, giving a usable bandwidth of 125MHz 
(equivalent to $\leq 0.1$\,km\,s$^{-1}$ at the frequencies 
observed).

Most of the pointed observations were made using 
position switching. Frequency switching was used only for some of the
HCO$^{+}$ and H$^{13}$CO$^+$ observations, and to check for emission-free
reference positions during position switching. Beam-switching was used
only occasionally, for sources which were known to have very compact emission.
Comparison of identical observations using each of these observing modes gives
excellent agreement in each case -- see Matthews (1996) for discussion
of observing modes.
Both grid mapping and raster mapping modes were used to make spectral
line maps. In some of the maps, the noise is greater towards the edge of the
map than at the centre, due to a reduced integration time per map
point at these positions. In the individual raster maps, integration
times of 4--6 seconds per
cell were used, as recommended by the JCMT Users Guide (Matthews 1996). 
Longer integration times per point were obtained by repeating the raster. 
Data from the JCMT data archive were also used to supplement our
observations where possible, as indicated in Table~4.

Data reduction was carried out using the {\small SPECX} package 
(Padman 1990). Baseline calibration was performed by subtracting a
fit through sections of baseline placed on either side of the
spectral feature of interest, carefully chosen to avoid suppression of low
level wing emission. Frequency switching tends to produce curved
or sinusoidal baselines. Curved baselines were removed by subtracting a
second order polynomial fit in the vicinity of the line.
This procedure was checked on several occasions by
making observations of the same object in both frequency-switched and
position-switched mode, giving very good agreement between the line profiles.

The spectra are calibrated in terms of 
the main beam temperature scale, apart from the maps, 
where the corrected antenna temperature, $T_{A}^{*}$, is used. 
The integrated intensity and centroid velocity, position-velocity and channel 
maps were also produced using the standard 
{\small SPECX} map-making facilities. To improve the appearance of the
contouring in these maps, the observed grids were first interpolated onto a 
grid with twice the spatial sampling rate.

The relation between centroid velocity $v_c$ over a given velocity interval in terms of $v_i$ and $T_{A_i}^*$, the velocity and antenna temperature 
of the $i$'th 
velocity channel, and the sum over the $n_{\rm ch}$ velocity 
channels lying in the 
specified velocity interval 
(Narayanan, Walker \& Buckley, 1998) and $\sigma_{T_A}$ the rms noise in a single
channel, assumed to be constant for all the channels in the velocity window, is given by:

\[ \frac{\sigma_{v_c}^2}{v_c^2}=\frac{n_{\rm ch}\sigma_{T_A}^2}{(\sum_i 
T_{A_i}^*)^2} +
\frac{\sigma_{T_A}^2\sum_i v_i^2}{(\sum_i v_i T_{A_i}^*)^2} 
- \frac{2\sigma_{T_A}^2 \sum_i v_i}{(\sum_i T_{A_i}^*)\cdot 
(\sum_i v_i T_{A_i}^*)}. \]

\noindent
This equation is used to derive the error bars for the centroid velocity
measurements presented below.

\section{NGC1333--IRAS2}

\subsection{Previous Observations}

NGC 1333 is a reflection nebula associated with the L1450 dark cloud,
which lies at a distance of $\sim$220pc (\u{C}ernis 1990), in the
Perseus molecular cloud complex (Sargent 1979). To the south of the 
reflection
nebula lies a $\sim 450$\,M$_{\odot}$ 
molecular core, with a central density and 
temperature of $10^4$cm$^{-3}$ and $\sim 18$K 
respectively (Warin et al., 1996; Lada et al., 1974). 
This is a site of highly active low and intermediate 
mass star formation, evidenced by the large concentration 
of T-Tauri and Herbig Ae/Be stars
(Lada, Alves \& Lada 1996; Aspin, Sandell \& Russell 1994), Herbig-Haro 
objects (Bally, Devine \& Reipurth 1996), and bipolar jets and outflows 
(Hodapp \& Ladd 1995; Liseau, Sandell \& Knee 1988) in and 
around the core. 
Loren (1976) originally suggested that the star formation 
in NGC1333 is being 
triggered by the collision of two dense molecular clouds. 
More recently, Warin 
et al. (1996) have argued that the morphology of the region 
supports a sequential star formation scenario, 
where outflows produced by one generation of stars compress
the surrounding gas,
which triggers collapse and further star formation.

\begin{figure*}
\setlength{\unitlength}{1mm}
\begin{picture}(60,90)
\includegraphics{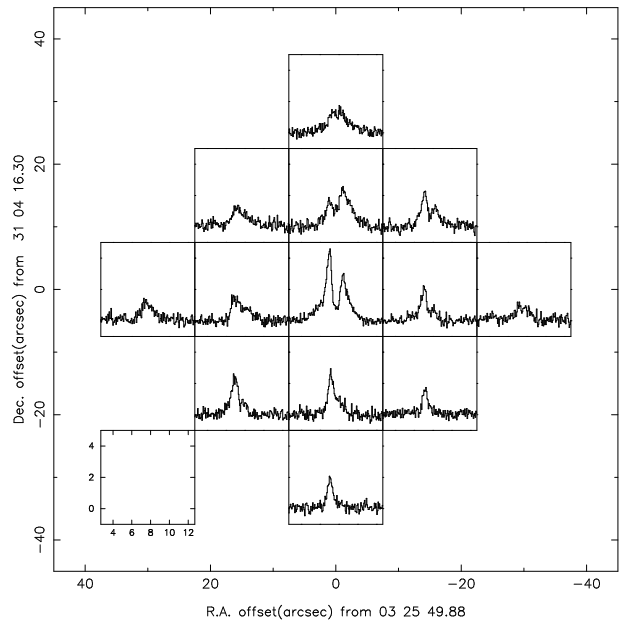}
\end{picture}
\caption{Grid map of $T_A^*$ for NGC1333--IRAS2 in the 
HCO$^+(J=4\rightarrow 3)$ transition in units of K.} 
\label{iras2hcogrid}
\end{figure*} 

Jennings et al. (1987) listed nine compact IRAS sources associated with
the NGC1333 cloud, five of which drive 
bipolar molecular outflows (Liseau et al. 1988). 
Of these, IRAS 2 and IRAS 4 have no associated
on-source optical or near-infrared emission. A full 
submillimetre continuum survey of 
NGC 1333--IRAS 2 was carried out by Sandell et al. (1994) using the JCMT. 
A strong compact peak was found at all wavelengths, with low-level 
emission at 800$\mu$m extending northwest 
and southeast of the peak, in a `flattened bar' morphology. 
Sandell et al. (1994) derived an envelope 
mass of $\sim 0.8$\,M$_{\odot}$ for the compact submillimetre source. 
The total far infrared luminosity of IRAS2 was estimated to be 17\lsun 
by Jennings et al. (1987) for a distance of 220pc. An upper limit 
to its total luminosity of 26\lsun was found by 
Ward-Thompson et al. (1996), who
quoted the ratio of bolometric to submillimetre luminosity to be 
$L_{\rm BOL}/L_{\rm SUBMM}\le 130$, below
the Class 0 threshold of 200 (Andr\'e et al., 1993).

Two bipolar outflows have been associated with IRAS2, 
observed in CO $(J=3\rightarrow 2)$ emission 
(Sandell et al. 1994), several transitions of CS 
(Langer, Castets \& Lefloch 1996; Sandell et al. 1994), 
and 2.12$\mu$m shock-excited molecular hydrogen emission 
(Hodapp \& Ladd 1995). 
The outflows are oriented nearly perpendicular to each other
on the plane of the sky. However, since they typically only affect the extreme line wings, they do not alter the results of our modelling.

\begin{figure*}
\setlength{\unitlength}{1mm}
\begin{picture}(60,90)
\includegraphics{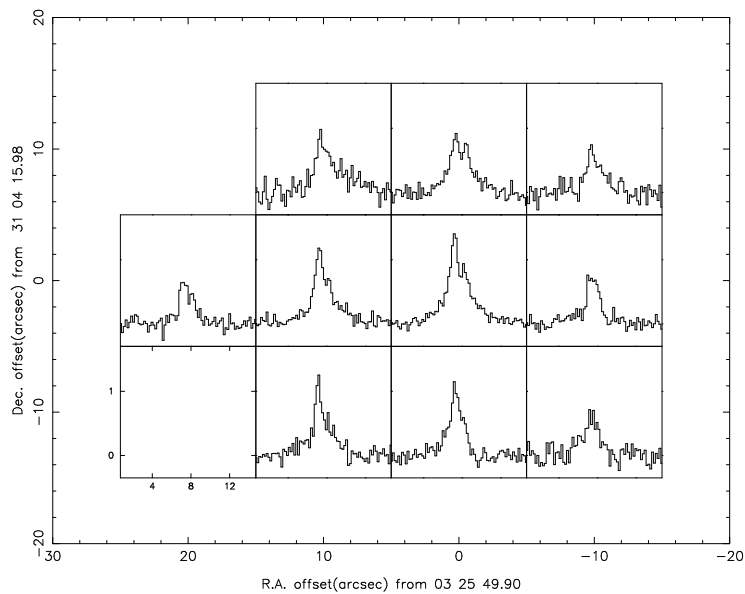}
\end{picture}
\caption{Grid map of $T_A^*$ for NGC1333--IRAS2 in the 
CS$(J=5\rightarrow 4)$ transition in units of K.}
\label{iras2csgrid}
\end{figure*} 

\subsection{New data}

Figure 19 presents 
spectra taken at the position of NGC1333--IRAS2 (hereafter IRAS2).
Both the \hcoft and \csff spectra show asymmetric
double peaked line profiles, with the blue peak stronger. The absorption dip
in the \hcoft line profile is particularly strong, and coincides in velocity
with the peak of the 
rarer isotopomer lines. Also visible in the main line \hco and CS 
spectra
are broad, low-level wings, which are probably tracing the outflowing 
gas associated
with this source. The peak of the \ceoto line is shifted towards the 
blue relative
to the \htcoft peak. Since \htcoft is expected to be the optically 
thinner of the
two lines, we interpret this as showing that the C$^{18}$O line is 
partially
optically thick in this object. 

Several properties of the spectra 
agree with the predicted qualitative signatures of infall in
protostellar envelopes (e.g. Myers et al. 1995; Zhou et al. 1993).
For optically thin lines, radiative transfer 
models predict a symmetric single peaked line 
profile, centred on the systemic velocity. 
Such a profile is shown by the \htcoft line, 
suggesting a systemic velocity of 7.7\kms . 
The asymmetric double-peaked line profiles skewed 
towards blue velocities seen in both the \hcoft 
and \csff spectra are predicted 
by radiative transfer models of infalling envelopes 
for optically thick lines.
The velocity of the self-absorption minimum may coincide with the 
systemic velocity,
or be red-shifted with respect to it, 
depending on the optical depth of the line
and the velocity and density structure of the envelope. 

The \ceoto line profile, which is slightly asymmetric, and blue-shifted
with respect to the systemic velocity, is also consistent with infall model 
predictions for a line with small but non-negligible optical depth. The only 
disagreements with the qualitative 
expectations of a pure infall model are the
strong high velocity wings seen in several of the spectra,
which we ascribe to the outflow (infall models do 
predict high velocity wings, but at 
a lower level than observed here), and the very small red-shift
of the peak of the \csss line profile, with respect to the inferred systemic 
velocity. 

Mardones et al. (1997) obtained on-source spectra of IRAS2 in 
CS($J=2\rightarrow 1$), N$_2$H$^+(J_{F_1F_2}=1_{01}\rightarrow 1_{02})$
and H$_2$CO $(J_{K_{-1}K_1}=2_{12}\rightarrow 1_{11})$, using the 
IRAM 30-m telescope. 
The CS line is single-peaked and approximately symmetrical, 
peaking at the systemic 
velocity (7.7\kms), and the N$_2$H$^+$ line has an infall type 
asymmetrical profile. 
Within the infall scenario, this would suggest that the CS line 
is more optically thin than the N$_2$H$^+$ line,
contrary to expectations. The H$_2$CO line profile observed by 
Mardones et al. (1997) has a 
complicated, multiple-peaked structure, with a very broad 
red-shifted wing, suggesting that this 
line is sensitive to the outflow. The line core is symmetrical 
and skewed to the red, which contrasts with
the blue-skewed profiles shown in Figure~20. 
The complete set of spectral line
observations at the position of this object 
do not, therefore, paint a wholly consistent picture.
However, they are in reasonable 
agreement with the qualitative expections for an infalling envelope,
despite
the fact that outflow emission is probably contributing significantly
to the line centres in many of the observed transitions. 

The spatial dependence of the observed line profiles may 
be able to shed light on
this question. Grid maps of IRAS2 in \hcoft and CS($J=5\rightarrow 4$) 
emission are presented in Figures~20 and 21 respectively. In both maps the 
emission is centrally peaked, and the strength of the self-absorption feature 
decreases away from the peak. The 
off-source spectra in the maps show an interesting 
variety of profiles. In the \hco map there is a clear trend towards broader 
linewidths from south to north, which is also discernable in the CS map. 
The CS map spectra show broader and relatively 
stronger wing emission than the 
\hco spectra, and may therefore be more sensitive 
to outflowing gas. In a detailed submillimetre 
spectral line study of the NGC1333--IRAS4 proto-binary 
source, Blake et al. (1995)
came to a similar conclusion, finding that the CS 
abundance is considerably enhanced
in the outflow, and \hco appears to be less depleted 
than neutral species in the dense protostellar envelope.

\begin{figure*}
\setlength{\unitlength}{1mm}
\begin{picture}(60,130)
\includegraphics{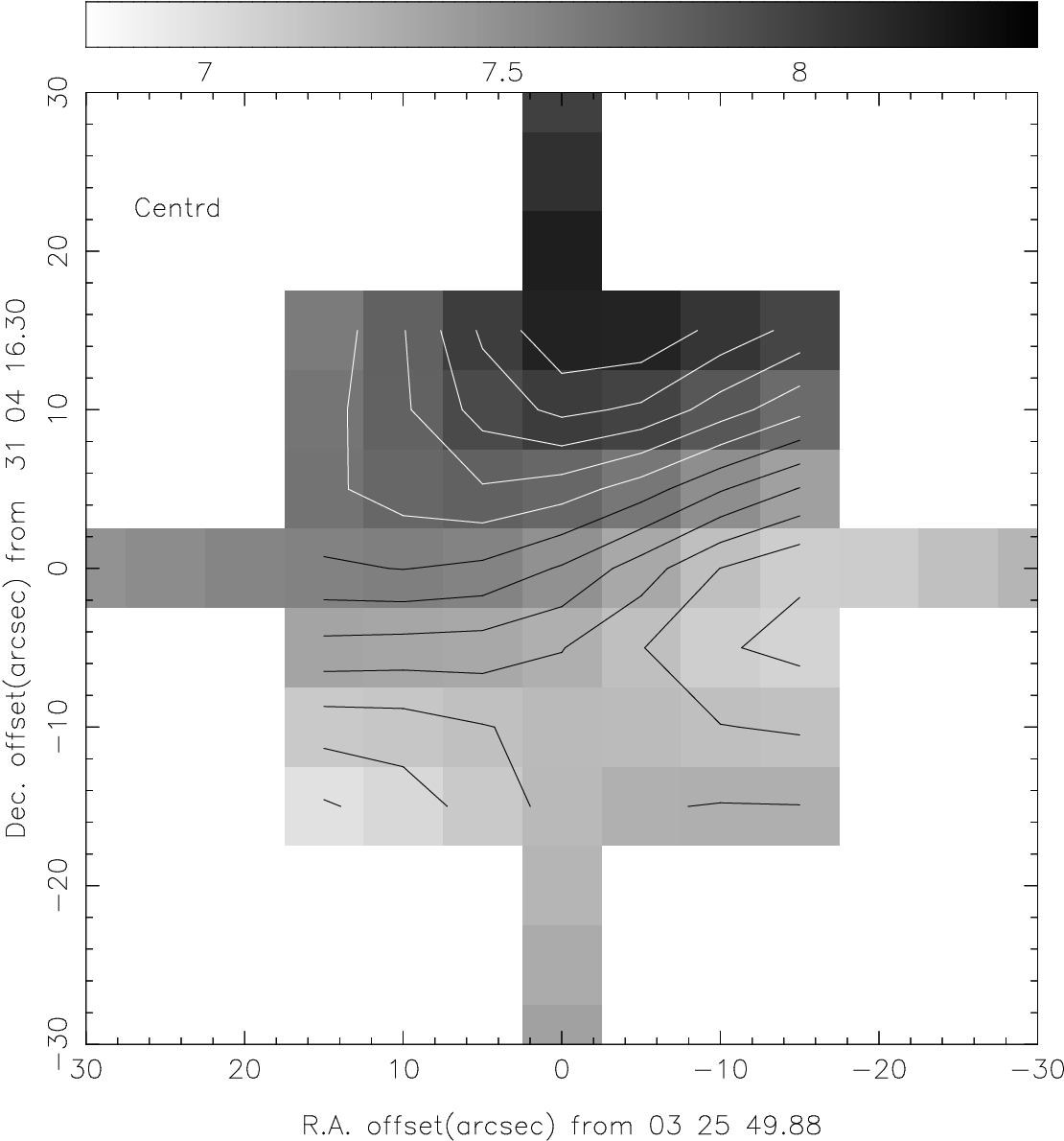}
\includegraphics{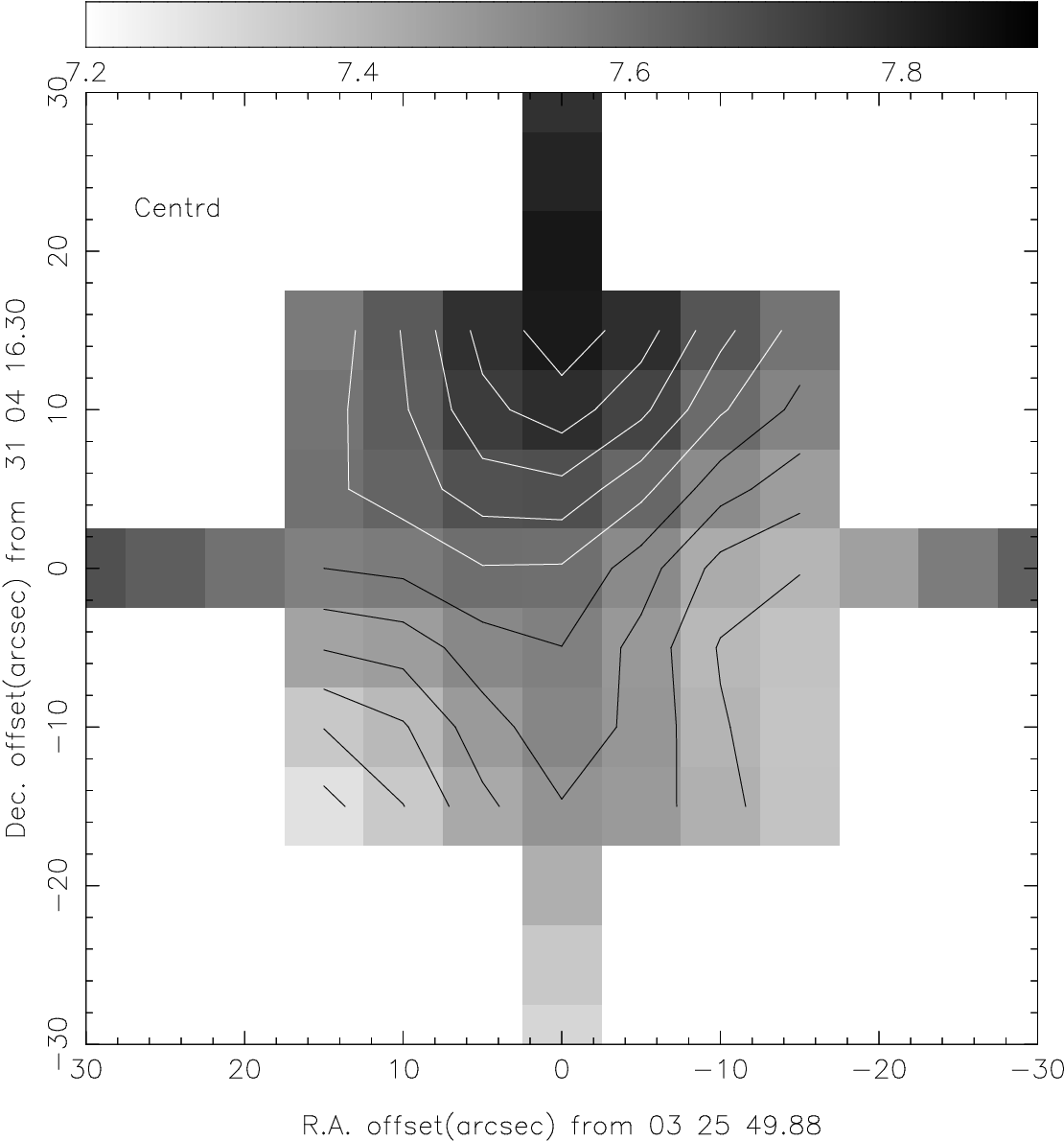}
\includegraphics{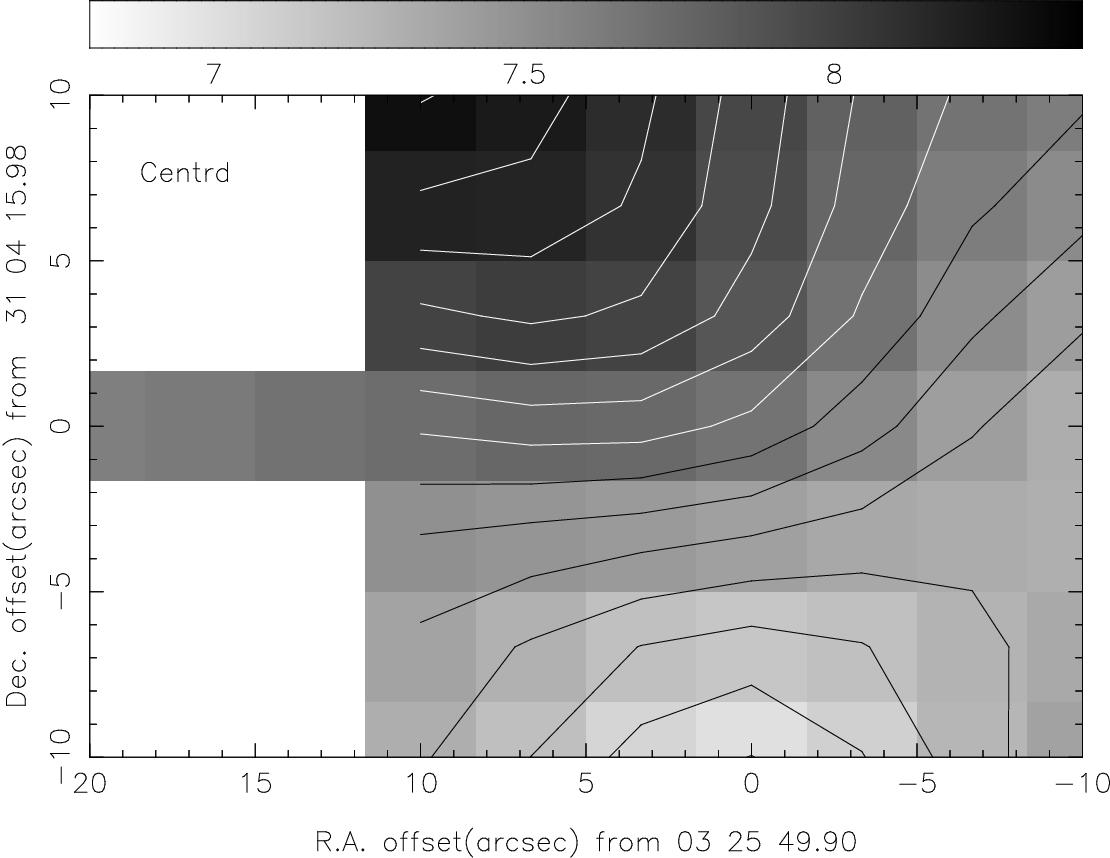}
\includegraphics{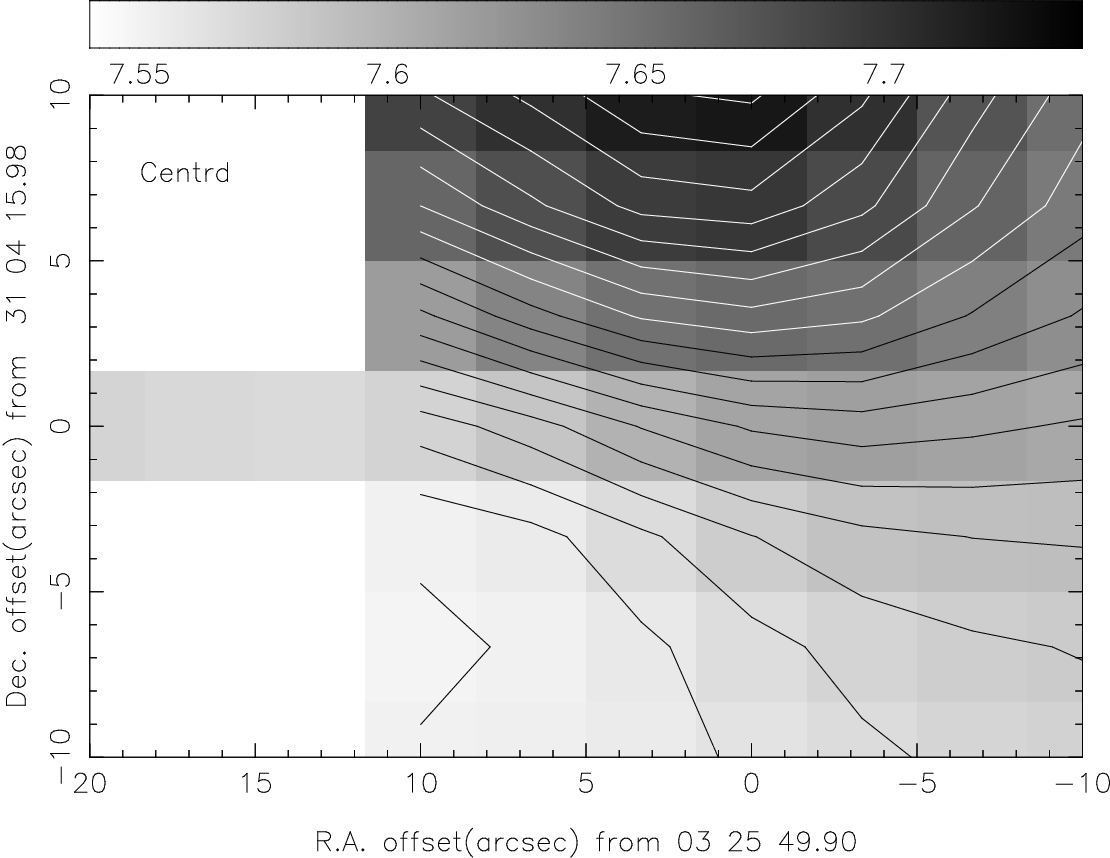}
\end{picture}
\caption{Comparison of centroid velocity maps of NGC1333--IRAS2
calculated over the velocity
range 3.7--11.7\kms (left) and 6.2--9.2\kms 
(right), to cover the whole 
line profiles and the line centres respectively. The 
top and bottom panels show \hcoft and \csff 
centroid velocity plots respectively. 
The greyscaling and contour spacing (0.05\kms ) 
are identical over all of the maps. 
The first light contour in each case marks a centroid velocity 
of 7.7\kms .} 
\label{iras2newcv} 
\end{figure*}

North of the map centre, a dramatic reversal of the 
line asymmetry in the \hco map is seen, 
and the CS spectra show weaker blue-asymmetries. 
One possible explanation for this is that
the emission from the northern (red-shifted) outflow 
lobe is enhancing the redshifted emission, perhaps due 
to an interaction with 
ambient gas at this position.

Centroid velocity maps calculated from the \hco and CS grid maps are 
shown in Figure~22. The interpretation of centroid 
velocity maps
requires some care, since the centroid velocity may be quite 
sensitive to the velocity range
over which it is calculated. To illustrate this, we show \hcoft and \csff 
centroid velocity plots calculated for two different velocity 
windows, one covering the 
entire line profile, including the high velocity wings (3.7--11.7\kms ), 
and one covering mainly the line centre (6.2--9.2\kms ). All of 
the plots show a clear 
velocity gradient along an axis with a position angle between 
0$^{\circ}$ and 
20$^{\circ}$. The velocity gradient is in the 
same sense as the north-south bipolar outflow 
suggesting a possible physical connection with it. 
The east-west outflow may be responsible 
for the small east-west velocity gradient apparent 
in the centroid velocity 
plots calculated over the wider velocity window. 

The `wide window' centroid velocity plots 
clearly show a much stronger velocity 
gradient than the `narrow window' plots, 
particularly in the \csff maps. This is consistent
with the idea that this gradient is caused by the 
north-south bipolar outflow, and also supports the 
suggestion that CS is a more sensitive outflow tracer 
than \hco . The agreement in the direction of the 
centroid velocity gradient
for both the narrow and wide velocity windows suggests 
that the outflow even affects the line profile shapes in
the line centres. This adds a serious complication to the 
interpretation of the line centre profiles in terms of 
infall models, since it is then extremely difficult to 
separate reliably the contribution of the envelope and outflow 
to the line profile. 

We now examine a possible alternative explanation for the velocity 
gradient seen in the centroid velocity maps, and the reversal of 
line asymmetry seen in the \hcoft map --- rotation of the 
protostellar envelope. Rotation will tend to produce a 
centroid velocity gradient across the envelope, and radiative transfer 
models which include both rotation and gas infall predict reversals of 
the infall asymmetry at certain off-centre positions 
(e.g. Zhou 1995), although the degree 
of reversal in the \hco spectrum north of the map 
centre in Figure~21 is larger than 
typically predicted by such 
models. Since the line wings of our CS and \hco 
observations are clearly affected by the outflow, we concentrate
on centroid velocities calculated over the line centre only.

\begin{figure*}
\setlength{\unitlength}{1mm}
\begin{picture}(60,75)
\includegraphics{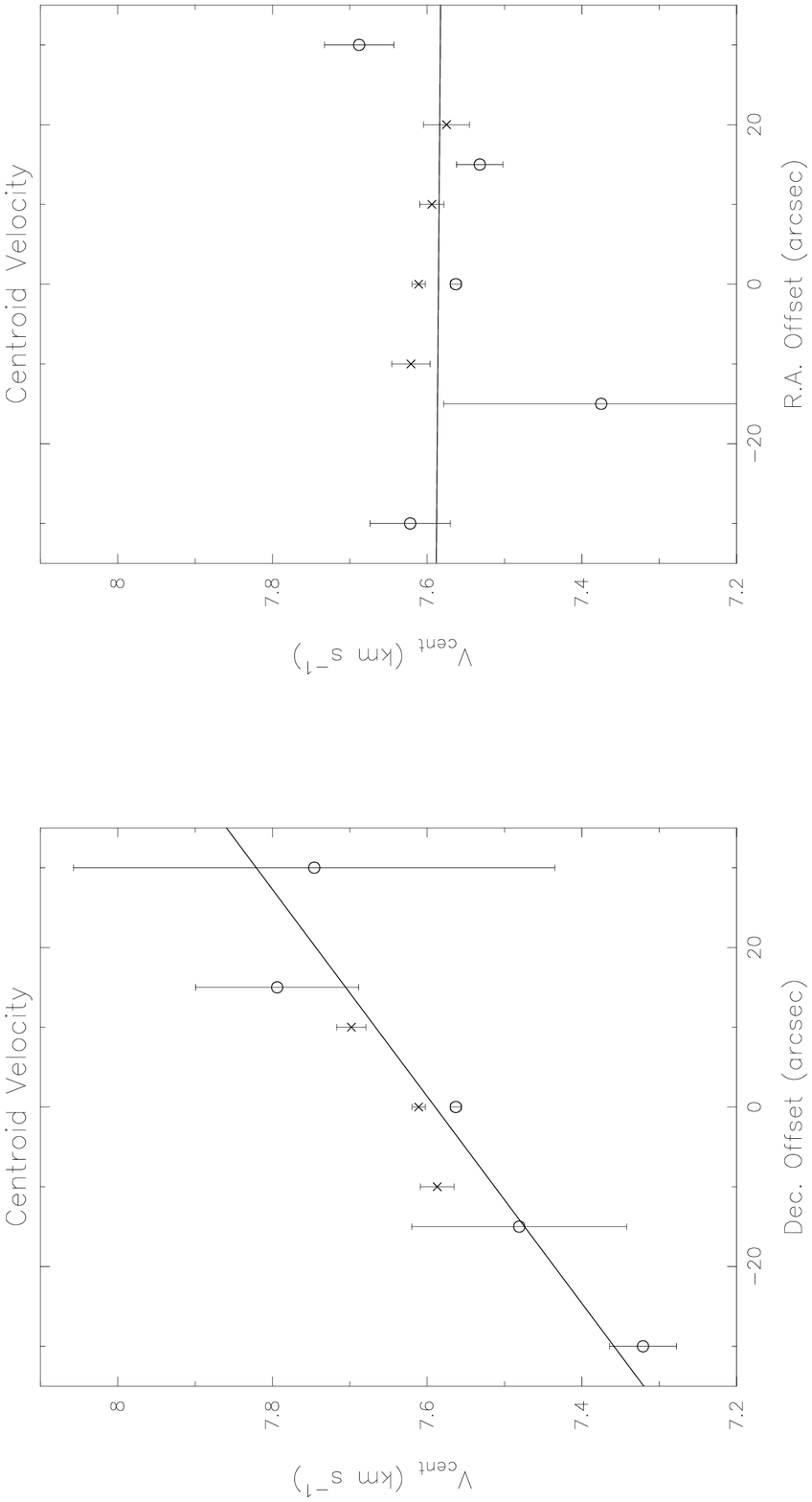}
\end{picture}
\caption{Plot showing weighted least-squares linear fits to the 
centroid velocities
along north-south (left) and east-west (right) axes through the position of 
the NGC1333--IRAS2 submillimetre continuum peak (at the map origin). 
The circles and
crosses denote \hcoft and \csff measurements respectively. The 
centroid velocity 
was calculated over the line centre (6.7--8.7\kms ).}
\label{iras2cvplot}
\end{figure*}

Figure~23 shows linear fits to 
the combined \csff and \hcoft centroid velocities 
calculated over the velocity window 6.7--8.7\,km\,s$^{-1}$, 
along north-south and east-west axes through the source.
The fit to the north-south centroid velocity gradient 
is $7.7(\pm 0.9)\times 10^{-3}$\kms arcsec$^{-1}$. For a distance of 220pc,
this is equivalent to 7.2$\pm$0.8\kms pc$^{-1}$. The 
corresponding numbers for the 
east-west velocity gradient are $-8.0(\pm 0.8)\times$ 
10$^{-5}$\kms arcsec$^{-1}$ 
and $7.5(\pm 0.7)\times 10^{-2}$\kms pc$^{-1}$. 
The interpretation of centroid velocities is 
complicated by optical depth effects, 
since both the transitions are optically thick in the line centre. 
Furthermore, infall motions in the envelope will 
tend to skew the profiles of
optically thick lines towards bluer velocities.

One indication of this is the fact that the centroid 
velocities along the east-west axis
are blue-shifted by $\simeq 0.1$\kms , on average, with 
respect to the systemic velocity. 
The measured centroid velocity gradient for an optically 
thick line is therefore an unreliable
measure of the actual velocity gradient in the object, 
and a proper treatment requires
full 3-D radiative transfer modelling, 
or the use of optically thin lines to
eliminate optical depth effects. However, for the purposes of 
the present discussion we assume 
that our centroid velocity gradient measurements give 
a reasonable indication of the actual line-of-sight 
velocity gradient across the object due to rotation.

Using this assumption a lower limit for the binding mass can be 
estimated by hypothesising that the rotation is centrifugally 
balanced by gravity, and that the rotation axis lies in the 
plane of the sky. This gives a lower limit,
because the actual rotation rate may be greater if the rotation 
axis is inclined out of the
plane of the sky, and magnetic fields and pressure gradients may 
also contribute to the support of
the envelope. Furthermore, if part of the envelope is infalling, 
as some of the observations suggest, 
then at some level the gravitational forces must dominate over the 
centrifugal support.

Assuming spherical symmetry, the mass $M$ interior to a radius 
$r$ is then given by:

\begin{equation}
M = \frac{rv(r)^2}{G\cos i} = \frac{r^3\Omega(r)^2}{G\cos i}, \label{mcent}
\end{equation}

\noindent
where $r$ is the distance from the centre of rotation, $v(r)$ is the 
line of sight
velocity at $r$ (relative to the systemic velocity), $\Omega(r)$ is
the angular velocity of the rotation at $r$, $G$ is the 
gravitational constant, and $i$ is the inclination angle of the 
rotation axis
from the plane of the sky. Using Equation~\ref{mcent} with
$i=0$, we find a lower limit of 0.4\msolar for the enclosed 
mass inside a 30 arcsec (6600AU) radius. Sandell et al. (1994) 
estimated a total dust and gas mass in the IRAS2 protostellar envelope of 
0.79\msolar, and to this must be added the mass of the central protostar. 

This analysis therefore does not
contradict the hypothesis that the line-centre 
centroid velocity gradient is tracing rotation about an east-west axis 
projected onto the plane of the sky. If this were correct, the most likely 
model would be that the east-west bipolar outflow is driven by the Class 0 
source, and the protostellar envelope is rotating about this outflow axis. 
The older north-south outflow would then have to be
driven by a separate, more evolved, and as yet unobserved object.
This seems unlikely, and we consider it more likely that the velocity 
gradient is tracing the north-south outflow rather than rotation.

\begin{figure*}
\setlength{\unitlength}{1mm}
\begin{picture}(60,90)
\includegraphics{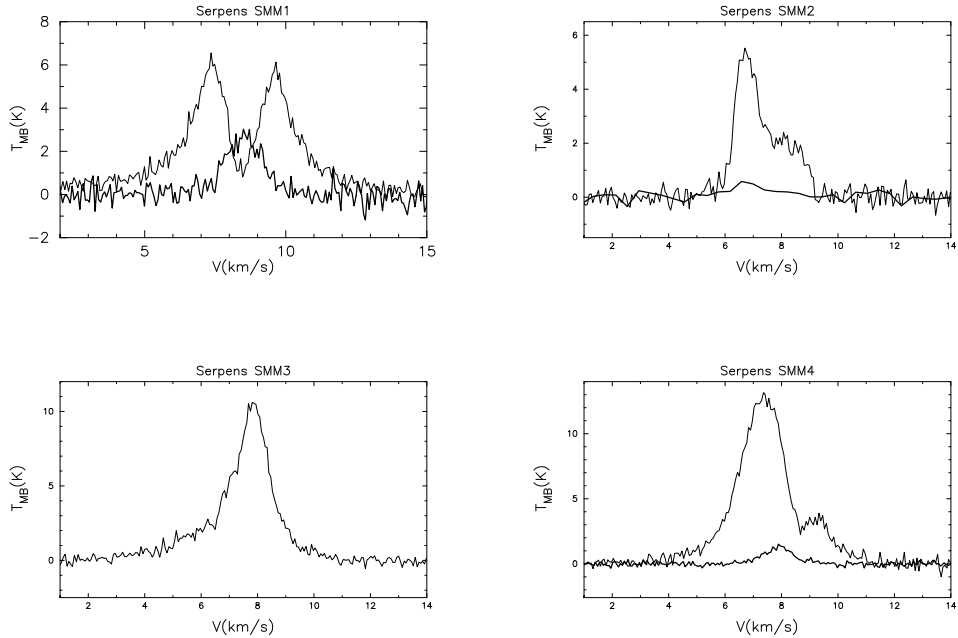}
\end{picture}
\caption{HCO$^+(J=4\rightarrow 3)$ and H$^{13}$CO$^+(J=4\rightarrow 3)$ 
(bold) spectra of the Serpens sources SMM1--4.} \label{serpenshco}
\end{figure*} 

\section{Serpens SMM1--4}

\subsection{Previous observations}

The Serpens molecular cloud is a low- to intermediate-mass 
star-forming region which has been the subject of intensive 
observational study (e.g. Eiroa 1991; Eiroa \& Casali 1992; 
White, Casali \& Eiroa 1995, hereafter WCE; Hogerheijde et al. 1998;
Testi \& Sargent 1998; Testi et al. 2000; Williams \& Myers 2000;
Davis et al. 2000). The distance to 
the region has been 
the source of some debate, with estimates ranging from 250 parsecs 
(Chavarria et al. 1988), to 750 parsecs (Zhang et al. 1988), based 
on spectroscopic 
parallaxes of stars believed to be physically related to the cloud. 
More recent studies (de Lara et al. 1991; Strai\u{z}ys, \u{C}ernis \& 
Barta\u{s}i\={u}t\.{e}, 1996) have identified mis-classifications of 
the spectral 
types of some of the stars used in the previous distance determinations, 
and suggest 
an actual distance of $300\pm 30$ parsecs.

Eiroa \& Casali (1992) carried out a comprehensive 
near-infrared survey of the region and positively identified 51 young stellar 
objects associated with the cloud. They estimated a lower limit of 
$\simeq 450$ pc$^{-3}$ for the average stellar density over a region of
$\simeq$0.5 parsecs diameter, and a star formation efficiency in the range 
8--28 per cent, suggesting that a bound 
cluster may be forming. The peak of the star 
formation activity is centred close to the Serpens reflection nebula (SRN), 
which is a diffuse optical and near-infrared nebula 
(G\'omez de Castro, Eiroa \& Lenzen 1988; Eiroa \& Casali 1992) 
powered by the luminous pre-main sequence star SVS2 (Strom, Vrba \& 
Strom 1976).

Maps in ammonia (Ungerechts \& G\"{u}sten 1984), C$^{17}$O (WCE),
CS (McMullin et al. 1994b) and far-infrared emission 
(Zhang et al. 1988; Hurt \& Barsony 1996) show the gas and 
dust in the central region to be concentrated in two cores, 
aligned approximately northwest-southeast. Estimates of the 
gas density and temperature
in the cores lie in the range $10^4-10^5$cm$^{-3}$ and 20--30K 
respectively, with 
higher temperatures and densities towards some of the embedded sources.
The $\sim$70M$_{\odot}$ northwestern core contains the well 
studied far-infrared source 
Serpens FIRS1 (Harvey et al. 1984), which is associated with a 
highly collimated, high
velocity ($\sim 300$ km\,s$^{-1}$) radio continuum jet 
(Curiel et al. 1993; 1996), and a linear 
near-infrared feature, approximately aligned with the 
northwestern lobe of the radio jet 
(Eiroa \& Casali 1989). This source was labelled SMM1 
(Casali, Eiroa \& Duncan 1993, hereafter CED),
and is the strongest of the
submillimetre continuum sources in the cloud. 

The more massive southeastern core includes the reflection nebula, and 
contains a small cluster of compact submillimetre continuum sources 
(CED), two of which (SMM2 and SMM4) 
have no near-infrared counterparts. 
Barsony (1997) used a maximum correlation algorithm to enhance the 
spatial resolution of the IRAS maps of this region,
and derived upper limits for the far-infrared luminosities of the 
submillimetre 
sources. SMM1--4, and S68N (McMullin et al. 1994b) were identified 
as Class 0 sources.
Hurt, Barsony \& Wootten (1996) have observed each of these sources 
in the high density
molecular tracer H$_2$CO. 
In the H$_2$CO$(J_{K_{-1}K_1}=3_{03}\rightarrow 2_{02})$ transition, 
four out of the five sources showed asymmetrical line profiles 
suggestive of infall. 
The remaining source, SMM1, showed a symmetrical double-peaked 
profile. Gregersen
et al. (1997) included Serpens SMM1--4 in their \hco survey of 
Class 0 sources, 
and found strong infall signatures towards SMM2 and SMM4.

The CO $(J=2\rightarrow 1)$ outflow map of 
WCE shows possible bipolar outflow associations with SMM1 
(southeast-northwest) and 
SMM4 (approximately north-south), and perhaps
SMM2 (southeast-northwest). Establishing the presence or 
absence of outflows driven by the other 
Serpens objects is hindered by the considerable confusion 
from overlapping outflow lobes. 

\begin{figure*}
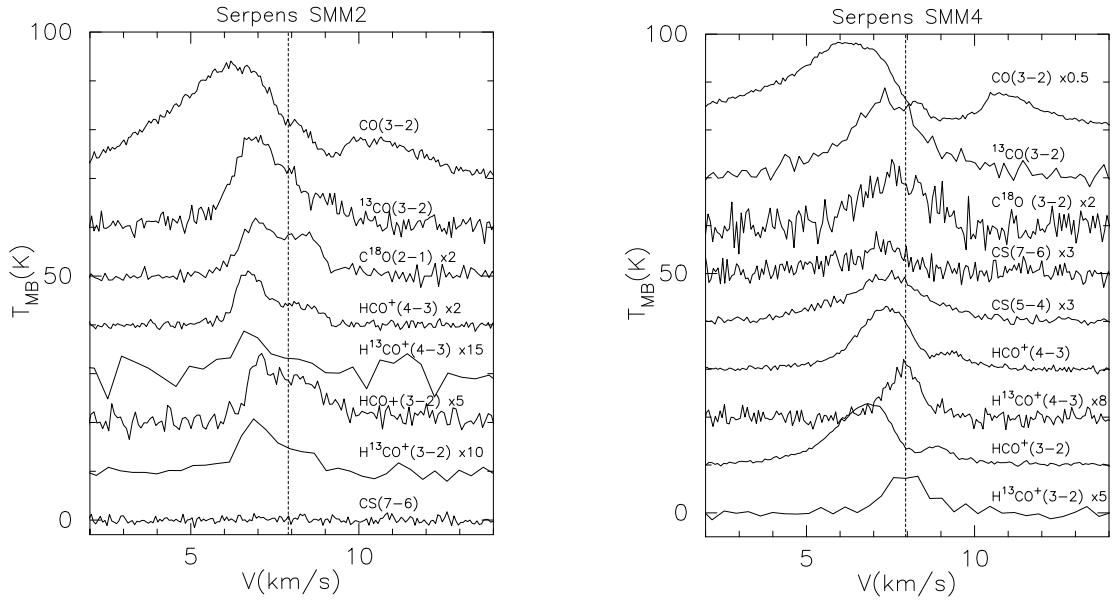

\setlength{\unitlength}{1mm}
\begin{picture}(60,85)
\includegraphics{fig25a.eps}
\includegraphics{fig25b.eps}
\end{picture}
\caption{(a) Spectra of various transitions
observed towards the position of the 
Serpens SMM2 continuum peak. The dashed line indicates
a velocity of 7.9\kms .
(b) Spectra observed towards the Serpens SMM4 continuum peak. 
The dashed vertical line indicates
a velocity of 7.95\kms.} \label{smm2-4spec}
\end{figure*} 

\begin{figure*}
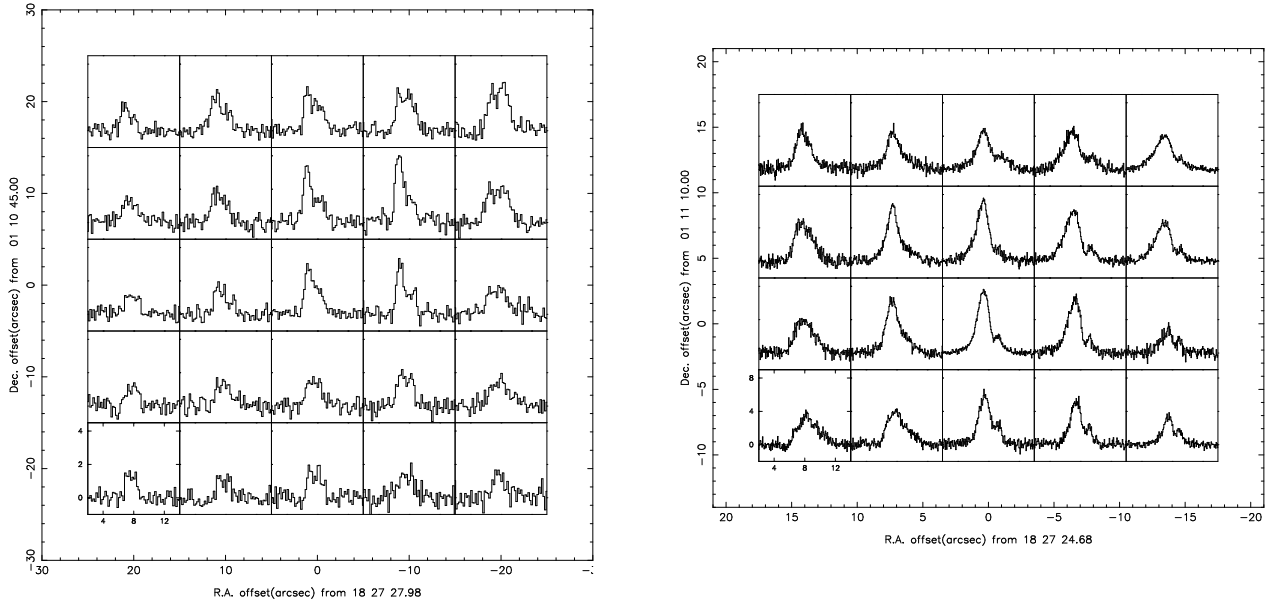

\setlength{\unitlength}{1mm}
\begin{picture}(60,85)
\includegraphics{fig26a.eps}
\includegraphics{fig26b.eps}
\end{picture}
\caption{(a) Grid map of Serpens SMM2 in the HCO$^+(J=4\rightarrow 3)$ 
transition (taken from the JCMT data archive).
The origin is at the position of the submillimetre continuum peak.
(b) Grid map of Serpens SMM4 in the HCO$^+(J=4\rightarrow 3)$ transition, 
centred on the submillimetre continuum peak.} 
\label{smm2-4hcogrid}
\end{figure*} 

\subsection{New data}

Figure~24 shows the \hcoft and \htcoft spectra 
observed towards the Serpens Class 0 
sources SMM\,1--4. Of the four sources, SMM2 and SMM4 show line 
profiles most suggestive of infall.
SMM4 in particular shows the classic signature of a double-peaked 
self-absorbed main line profile
skewed heavily towards the blue, with an absorption minimum redshifted 
relative to the 
peak of the optically thin isotopic line. The SMM1 main-line profile 
shows a deep absorption minimum, 
close to the peak velocity of the isotopic line, and a very mild blue 
asymmetry. Strong high 
velocity wings are also visible, which are probably tracing the energetic 
bipolar jet and outflow driven by this source. The \hcoft spectrum 
observed towards SMM3
does not show any sign of self-absorption, but a strong blue-shifted 
wing, probably tracing 
an outflow, is clearly seen. In the following we concentrate primarily on 
the sources SMM2 and SMM4, as the best protostellar infall candidates 
in the Serpens cloud.

\begin{figure*}
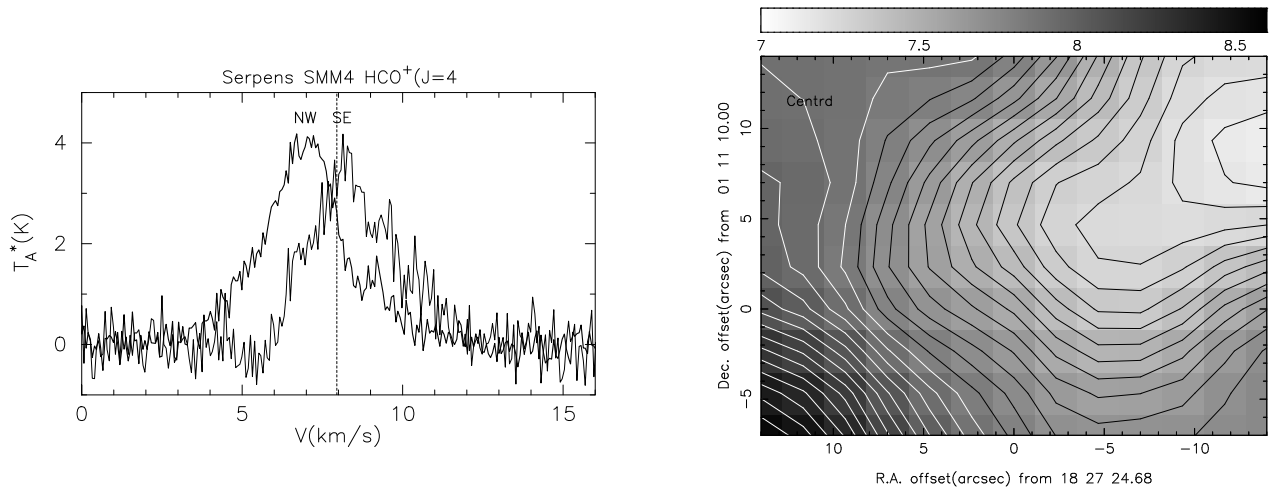

\setlength{\unitlength}{1mm}
\begin{picture}(60,75)
\includegraphics{fig27a.eps}
\includegraphics{fig27b.eps}
\end{picture}
\caption{(a) Comparison of spectra in the northwest ($-$14,+14) 
and southeast (+14,$-$7) corner 
of the \hcoft grid map of Serpens SMM4 shown in Figure~26(b).
(b) Centroid velocity map derived from the \hcoft grid map of 
Serpens SMM4, over the velocity range 5.0-11.0\kms . 
The contour spacing is 0.05\kms.}
\label{smm4velgradhcocv}
\end{figure*} 

Figure~25(a) shows
the on-source spectra observed towards Serpens SMM2. Blue
asymmetric line profiles are seen in most of the transitions, 
although \csss emission
was not detected above the $\sim$0.2K noise of the observation. 
There is a remarkable similarity
between line profiles of the main line \hco transitions and their 
isotopic counterparts, although the signal to noise ratio in the 
\htcoft line profile may be too low to attach much weight to this. 

Figure~26(a) shows a grid map of Serpens SMM2 in the \hcoft transition. 
This map was obtained from the
JCMT data archive, and the observing runs during which these data were 
taken are described in WCE. The \hcoft emission
is seen to peak northwest of the continuum peak.
However, the bright line profiles 
observed towards the
north-western corner of the map include a contribution from the outer 
envelope of SMM4,
which lies at an offset of ($-$25,+50), and shows extended bright emission 
in \hcoft -- see Figure~27. 
This may in part explain the similar velocity structure of the isotopic 
and main line profiles discussed above, if
the emission arises from overlapping clumps with different velocities 
along the line of sight. There is 
no reliable way to separate the overlapping contributions to the 
line profiles, and we therefore do not discuss this source any further.

The spectra observed towards the position 
of the Serpens SMM4 submillimetre continuum peak are shown in 
Figure~25(b). The optically thin 
lines -- \htcott and \htcoft -- peak at a velocity of 7.95\kms. The
\ceott line peaks at 7.7\kms, suggesting that 
there is a non-negligible optical depth in this line, as we have
found previously. There is 
good qualitative agreement between the line profiles shown and 
typical radiative transfer model predictions for an
infalling envelope (see the discussion above of the NGC1333--IRAS2 
on-source spectra). 

Both the main line \hco transitions show the asymmetric 
double-peaked line profile skewed towards blue velocities.
We also note that the 
\cott line also shows this type of profile. However, CO line profiles 
are not as robust infall indicators as HCO$+$, since they
suffer greater confusion from foreground material and outflowing gas. 
The isotopic 
\hco line profiles shown are also in qualitative agreement with model 
predictions, 
being single-peaked, and peaking between the velocity of
the absorption minimum and the blue-shifted peak of the main-line profile. 

Further 
qualitative agreement between model predictions and observations is shown 
by three 
isotopic CO lines. Models predict that as the optical depth of the observed 
transition increases, the line profiles progress from single peaked and 
symmetric, 
to single-peaked and skewed to the blue, to double-peaked with the blue 
peak brighter than the red (Myers et al., 1995). This is just the 
progression observed in the CO line profiles as the isotopic abundance 
increases. 
The CS line profiles represent the intermediate optical depth stage, 
being single 
peaked and skewed to the blue. 
Taken as a whole, therefore, the set of line profiles shown in
Figure~25(b) are remarkably consistent with the qualitative 
expections for an infalling envelope. 

Figure~26(b)
shows a grid map of the SMM4 envelope in the \hcoft transition. 
The strongest emission is found at the origin and at the
grid position immediately north of the origin. Blue-skewed line profiles 
are evident in most of the spectra shown, although double peaked line 
profiles are only found in the central and western
map columns. Comparison of the spectra in the southeast and northwest 
corners of the map reveals a shift in the peak
velocity of about 1\kms towards the blue along the southeast-northwest 
direction -- see Figure~27(a). 
A centroid velocity map calculated from this grid map is shown in 
Figure~27(b), from which a southeast-northwest centroid
velocity gradient of $\sim 30$\kms pc$^{-1}$ is measured. Also apparent in 
this plot is a `blue-bulge' encroaching onto
the red-shifted side of the velocity gradient from the blue-shifted side. 
As discussed above, this is a predicted signature of infall in the presence
of rotation. 

\begin{figure*}
\setlength{\unitlength}{1mm}
\begin{picture}(60,75)
\includegraphics{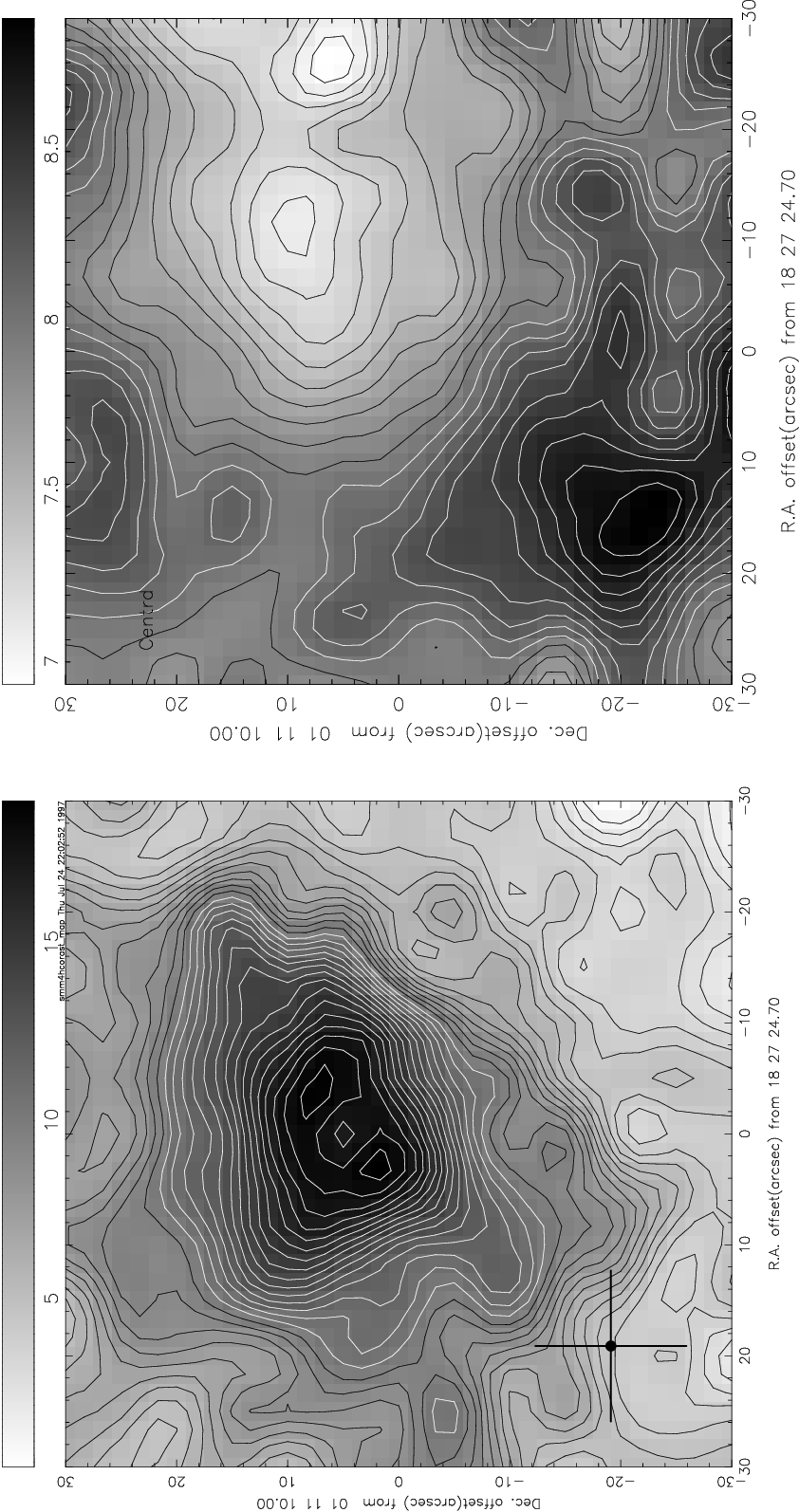}
\end{picture}
\caption{\hcoft integrated intensity (left) and centroid velocity 
(right) maps of Serpens SMM4, over the velocity interval 4--12\kms . 
The peak contour of the integrated intensity map is 18.6K\kms, and 
the contour 
spacing is 0.6K\kms ($\simeq 1\sigma$). In the centroid velocity map, 
the minimum and maximum contours levels are at 6.8 and 8.8\kms 
respectively, and 
the contour spacing is 0.1\kms. The rest velocity of the source 
($\simeq 7.95$\kms) lies in the transition interval between the 
dark and light 
contours. The cross indicates the effective FWHM beam size.}
\label{smm4hcogreycv}
\end{figure*} 

A larger area \hcoft raster map of SMM4 was also made, and is shown in 
the left hand plot of Figure~28 as an integrated 
intensity map. The \hcoft
emission is seen to be strongly enhanced in the region of SMM4 and is well
resolved in our beam (the transition 
from dark to light contours in the figure 
marks the half-maximum level). 
The emission is approximately centred around the
offset position (0,+5) on the map. Towards this point the integrated 
intensity flattens off into a plateau, 
and there is a shallow minimum at the
central point itself, surrounded by horseshoe-shaped 
peak. The lower contours 
reveal an extension of the peak towards the northeast, 
and a flattened linear 
feature oriented northwest-southeast.

The right hand panel of Figure~28 shows the centroid velocity map 
calculated from the raster map.
The southeast-northwest velocity gradient is again very apparent.
Comparison of integrated intensity and centroid velocity maps shows
that the region of brightest integrated \hcoft emission coincides with 
the blue-shifted 
part of the velocity gradient, which makes the symmetric 
appearance of the 
plot about the line rest velocity all the more remarkable. 
The axis of the steepest velocity gradient (P.A.$\simeq -40^{\circ}
\pm 5^{\circ}$) is close to (but not
exactly coincident with) the axis of the elongated feature in the 
integrated intensity map (P.A.$\simeq -50^{\circ}\pm 3^{\circ}$). 

The steepest velocity gradient is 
$-4.1(\pm 0.15)\times 10^{-2}$\kms arcsec$^{-1}$, which implies a 
gradient of $28\pm 1$\kms pc$^{-1}$, for a distance of 300pc. 
The centroid velocities along this axis clearly shift towards bluer 
velocities at positions near to the central peak. This `blue-bulge' 
signature,
parallel to the rotation axis of a rotating, infalling protostellar 
envelope, was first
predicted by the radiative transfer analysis of Walker et al. (1994).

As before, we estimate the lower limit on the mass assuming the
observed velocity gradient is due to rotation and the rotation axis lies 
in the plane of the sky. We consider the mass enclosed within the
limits of the steep velocity gradient between position offsets of 
$-$20 and +10 arcsec from the \hco peak, and obtain a lower 
limit of $\sim 1.9$\msolar (for a distance of 300 parsecs). 
Hurt and Barsony 
(1996) estimated a gas mass of 3\msolar surrounding SMM4, 
derived from dust continuum
emission, which, along with the mass of any embedded protostars, is 
sufficient to provide the centripetal force required to bind 
the rotation, as long as
the rotation axis is close to the plane of the sky. Nevertheless, 
if our analysis is 
correct, centrifugal effects are likely to have a very 
significant influence
on the equilibrium and dynamics of the envelope gas. 
The elongation seen in the integrated \hcoft map,
which is aligned approximately perpendicular to 
the inferred rotation axis, may well in fact be
caused by centrifugal flattening of the outer envelope
(e.g. Pudritz \& Norman 1986). 

The alternative possible explanation for the
observed velocity gradient in the \hcoft maps could be
outflowing gas. Gregersen et al. (1997) 
detected a velocity gradient in the same sense and along the same axis 
in their \hcott map of SMM4, but interpreted it in terms of outflow
rather than rotation. SMM4 has previously been associated 
with a north-south bipolar outflow (WCE), but not aligned with
the gradient we observe. Examination of the channel maps in our own
data likewise reveals no evidence for an outflow in this direction.
Thus we believe our rotation hypothesis
to be the more likely explanation.

\begin{figure*}
\setlength{\unitlength}{1mm}
\begin{picture}(60,90)
\includegraphics{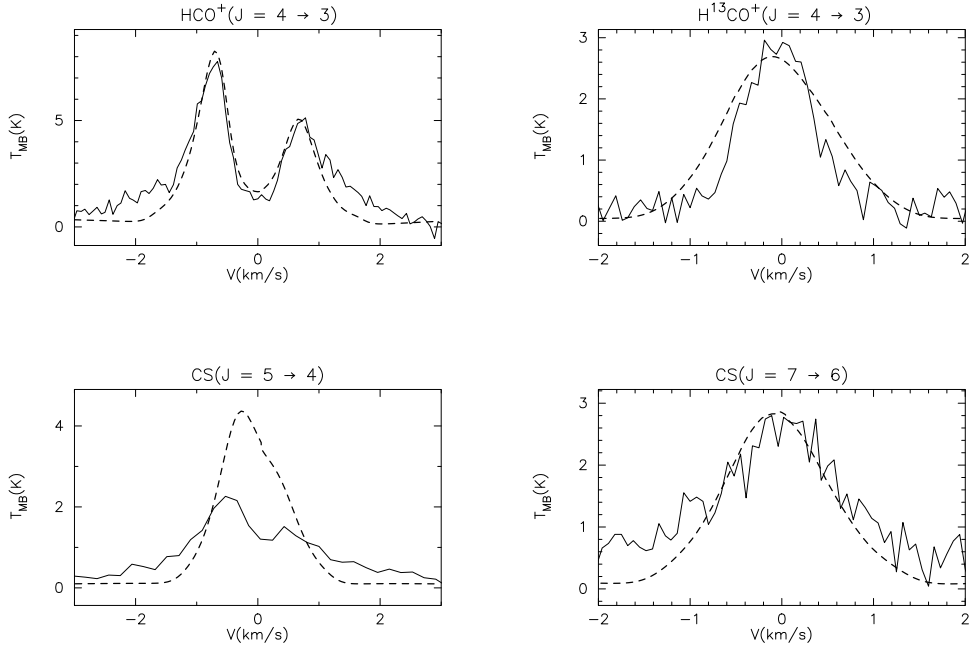}
\end{picture}
\caption{Model fits (dashed lines) to the on-source 
spectra (solid lines) observed towards NGC1333--IRAS2. 
The assumed systemic velocity is 7.75\kms, which has 
been subtracted from the 
velocities in the observed spectra (see text for details).} 
\label{iras2spec2}
\end{figure*}

\section{Modelling the data}

\subsection{NGC1333--IRAS2}

Figure~29 shows our best model fit to the \hco and CS line
profiles observed towards NGC1333-IRAS2. We approached the modelling by first 
searching for a fit to the \hcoft profile, choosing realistic physical 
parameters. We aimed to achieve consistency in our
model fits with previously determined properties of the objects, 
such as the envelope mass and 
ambient gas temperature. 
Once a suitable fit was found, the same model was used
to find the best fit to the \htcoft profile, but in 
this case the only free parameter 
was the [H$^{12}$CO$^{+}$]/[H$^{13}$CO$^+$] ratio, 
which we required to lie in the 
range 50--100. Depending on the quality of this fit, 
the model was either rejected, or 
tested further against the CS line profiles. 

We attached less weight to the CS profiles in the 
model fitting, since CS tends to be more affected by outflow emission 
than \hco (e.g. Blake et al. 1995). Langer et al. (1996) 
mapped the entire
NGC1333 core region in the \csff, $(J = 3 \rightarrow 2)$ and 
$(J = 2 \rightarrow 1)$ transitions
using the IRAM telescope, and found that the CS emission extended 
over a very wide area. The 
CS profiles may therefore be significantly affected by emission and 
absorption from the gas
in the foreground core. 

The parameters used in the model fit are listed in Table 5.
The outer radius of the cloud was set at 10\,000\,AU, and 
an innermost shell radius
of 15\,AU was used. A distance to the cloud of 220\,pc was 
assumed (\u{C}ernis 1990). The total envelope 
mass within the outer radius of the cloud inferred from the 
molecular hydrogen density 
profile is 3.7\,M$_{\odot}$.
The values of $r_{\rm inf}$ used in the infall velocity and density 
profiles were 6000\,AU and 900\,AU 
respectively, and 
$a_{\rm eff}=0.29$\kms was used in both cases. No acceptable 
fit could be found which used
identical values of $r_{\rm inf}$ for both profiles. In other words, 
we could not find a solution 
where the density and velocity profiles taken together are 
consistent with the singular isothermal sphere model. 

\begin{table}
\begin{center}
\begin{tabular}{ll} \hline
Parameter & Value \\ \hline
$v_r(r<6000\,{\rm AU})$ & 
$[-0.29\left(\frac{r}{6000\,{\rm AU}}\right)^{-\frac{1}{2}} $\\
& \hspace*{0.25cm}
$+0.29\left(\frac{r}{6000\,{\rm AU}}\right)^{0.15}]\ {\rm km\,s}^{-1}$\\
$v_r(r>6000\,{\rm AU})$ & $0\,{\rm km\,s}^{-1}$\\
$n_{{\rm H}_2}(r<900\,{\rm AU}) $ & 
$2.46\times 10^{6}[ 
0.35\left(\frac{r}{900\,{\rm AU}}\right)^{-\frac{3}{2}} $ \\
& \hspace*{0.25cm}
$+0.65\left(\frac{r}{900\,{\rm AU}}\right)^{-0.64} ]\ {\rm cm}^{-3}$\\
$n_{{\rm H}_2}(r>900\,{\rm AU})$ & $2.46\times 10^{6}
\left(\frac{r}{900\,{\rm AU}}\right)^{-2}\ {\rm cm}^{-3}$ \\
$T(r<2400\,{\rm AU})$ & 
$17\,\left(\frac{r}{2400{\rm AU}}\right)^{-0.4}\ {\rm K}$\\ 
$T(r>2400\,{\rm AU})$ & $17\,{\rm K}$\\ 
$\Delta v_{\rm tb}({\rm FWHM})$ & $0.60\,{\rm km\,s}^{-1}$\\ 
$X_{\rm HCO^+}$ & $6.0\times 10^{-9}$\\
$X_{\rm CS}$ & $6.0\times 10^{-9}$\\
$^{12}{\rm C}/^{13}{\rm C}$ & 90\\ 
\hline
\end{tabular}
\caption{The parameters used in the model of NGC1333--IRAS2.}
\end{center}
\end{table}

The fit to the \hcoft line profile is very good between $-$1 and +1\kms
(relative to the systemic velocity of 7.75\kms ). 
At larger velocities in both
the red and blue wings of the line, the model significantly 
underestimates the emission.
The most likely origin for the excess high velocity emission is from one or 
both of the bipolar outflows associated with this source -- the
clear break in the slope of the observed line profile at 
$\simeq\pm$1.3\kms 
also suggests a separate origin for the high velocity emission. We have
not tried to estimate how much the outflow contributes to the line profile
at lower velocities, so this is another source 
of uncertainty in the model. 

\begin{figure*}
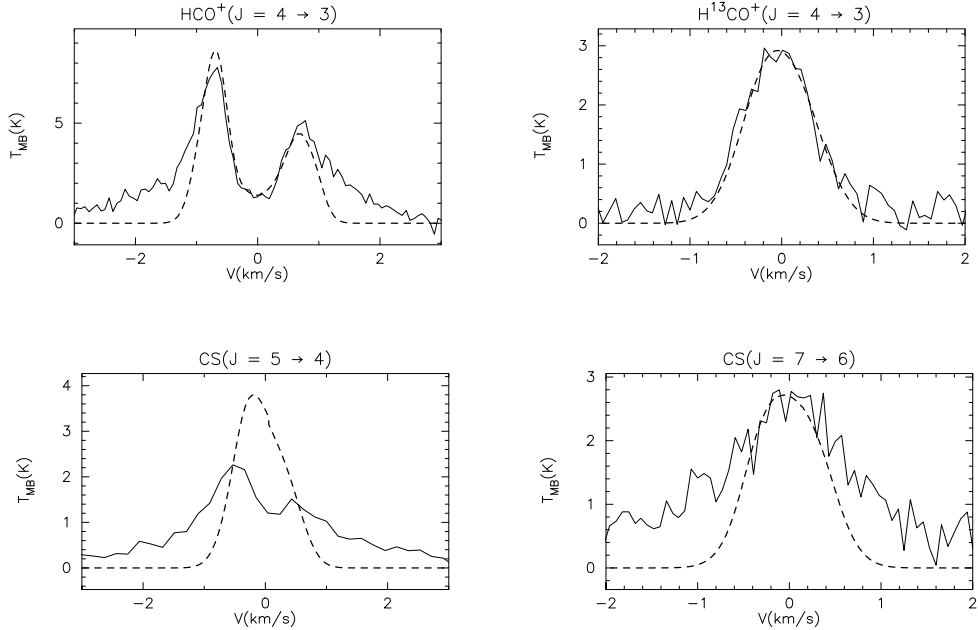

\setlength{\unitlength}{1mm}
\begin{picture}(60,90)
\includegraphics{fig30a.eps}
\includegraphics{fig30b.eps}
\end{picture}
\caption{Model fits using the same model parameters
as in Figure~30, but with the infall 
velocity set to zero inside a radius
of 800\,AU, and a factor of 4 reduction in the CS 
relative abundance.} 
\label{lowvcent}
\end{figure*}

The \htcoft model fit agrees well with the peak velocity and 
intensity in the observed line, although the model profile has
slightly overestimated the line width. The width of the
\htcoft line in the model is mostly determined by the systematic 
and turbulent
velocities near the centre of the cloud, since the warm dense gas 
in this region
has the strongest intrinsic emission, and for an optically thin 
tracer this 
emission is unattenuated by the outer envelope on its way to the 
observer. 
The asymmetry in the main line profile is mainly sensitive to the 
systematic and
turbulent velocities at larger radii. If the infall velocity at 
large radii is chosen to produce agreement with the asymmetry 
in the \hcoft profile, 
then the inferred velocity at small radii always produces too 
much broadening 
in the \htcoft line, as a result of the $v\propto r^{-1/2}$ velocity law.

The largest discrepancy between the observations and the model is 
in the \csff line,
where the model predicts a substantially higher intensity in the 
line core than 
is observed. Better fits to the overall intensity of the CS line 
(but not the line shape) could be found 
by reducing the CS relative abundance, although this then 
underestimated the intensity of the 
\csss line. This might be explained if
the CS relative abundance may vary with 
radius, but we did not pursue this. 
The value of the CS relative abundance 
used in the model fits was simply optimised to give the best 
agreement with the \csss line. 

Figure~30 shows an alternative fit to the \hcoft and 
\htcoft line profiles 
obtained by setting the infall velocity to zero inside a radius 
of 800\,AU, keeping all
other parameters the same. The main line profile is affected 
remarkably little by this 
dramatic alteration to the inner velocity field
(although the fit is slightly worse), which underlines 
the point that each observed 
transition constrains only a limited domain of parameter space in 
the model envelope. The
CS line profiles (in which the CS relative abundance has been 
lowered by a factor of 4 relative
to the original model fit) are still not well fitted by this model. 
In both Figures~30 \& 31
the predicted \htcoft and \csss line profiles are almost identical to 
each other, and we could not produce simultaneous fits to the velocity 
widths of both
observed profiles. Since CS emission shows a greater tendency for 
contamination by
the outflow, we are inclined to attribute the larger velocity width 
on the \csss line to the outflow, and therefore we tend to prefer
the former fit described above.

\begin{figure*}
\setlength{\unitlength}{1mm}
\begin{picture}(60,130)
\includegraphics{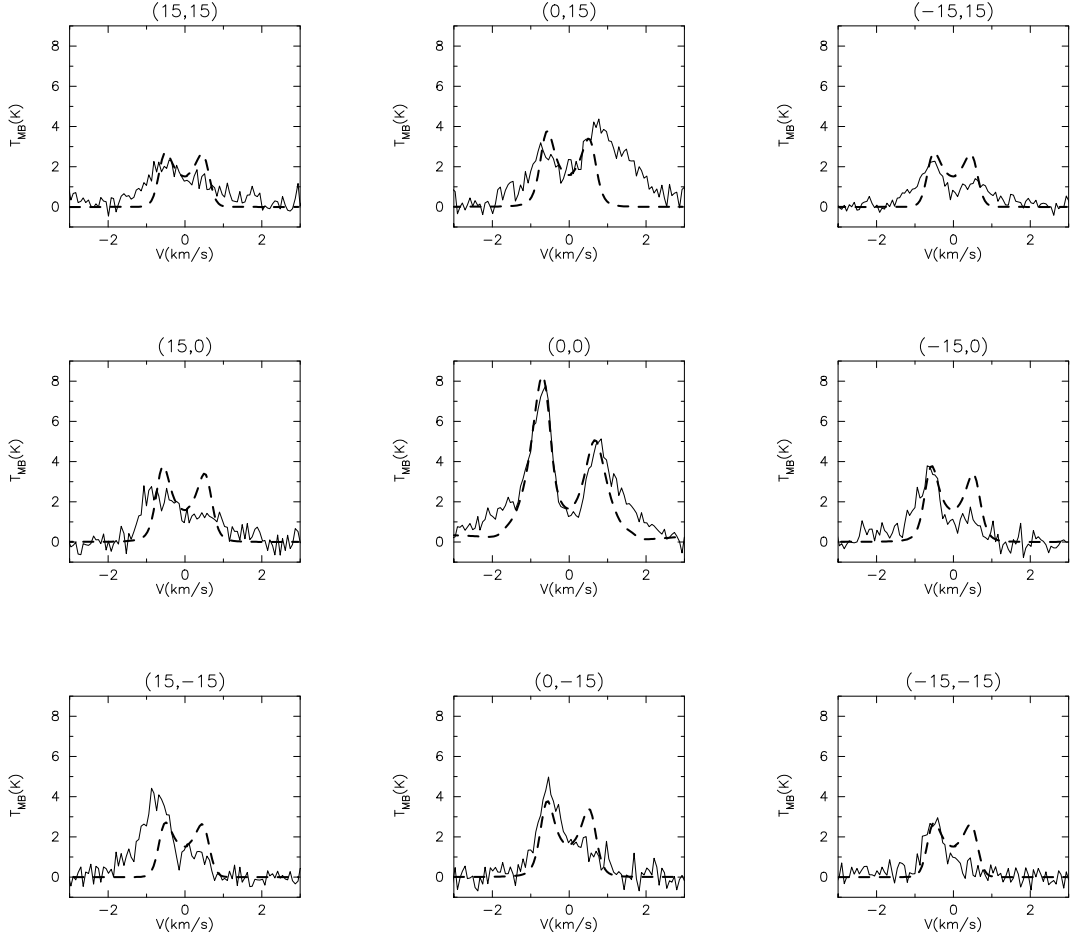}
\end{picture}
\caption{Comparison of the off-centre \hcoft 
line profiles observed towards NGC1333--IRAS2 
(solid lines) with the model predictions (dashed lines). 
The offsets (in arcsec) from the IRAS2 position are given 
in the title of each plot.}
\label{iras2hcomap}
\end{figure*}

In Figure~\ref{iras2hcomap}, we compare the off-centre \hcoft line profiles
with the model predictions. Because of the assumption of spherical 
symmetry, the
predicted line profiles depend only on the impact parameter of the line 
of sight relative to the 
origin, and not on its direction. The decrease in line intensity with 
increasing impact parameter is reasonably well matched by the model. 
However, the line shapes 
at the off-centre positions show less
good agreement. With the exception of the 
spectrum at the (0,+15) position, the off-source line profiles are 
all skewed towards the blue.
The model profiles are much more symmetric, as a result of the fall-off 
in infall
velocity at large radius, which is again a consequence of the 
$v\propto r^{-1/2}$ infall
velocity law. It is very difficult to accommodate the strong
high velocity emission seen in several of the off-source positions
-- most prominently at (0,+15) and (+15,$-$15) -- 
within a plausible infall model, particularly given the highly 
asymmetric distribution of the high velocity emission in both 
position and velocity, which cannot
be fitted by a spherically symmetric model.

We explored the possibility above
that the centroid velocity gradient measured
across the NGC1333--IRAS2 envelope using \hcoft 
and \csff spectral line maps could be explained
by rotation, and concluded that although rotation 
could not be ruled out, outflow was
the more likely explanation. On the basis of the 
results in Figure 31, we believe that
the north-south bipolar outflow, and not rotation, is responsible for the 
centroid velocity gradient. 

\begin{figure*}
\setlength{\unitlength}{1mm}
\begin{picture}(60,90)
\includegraphics{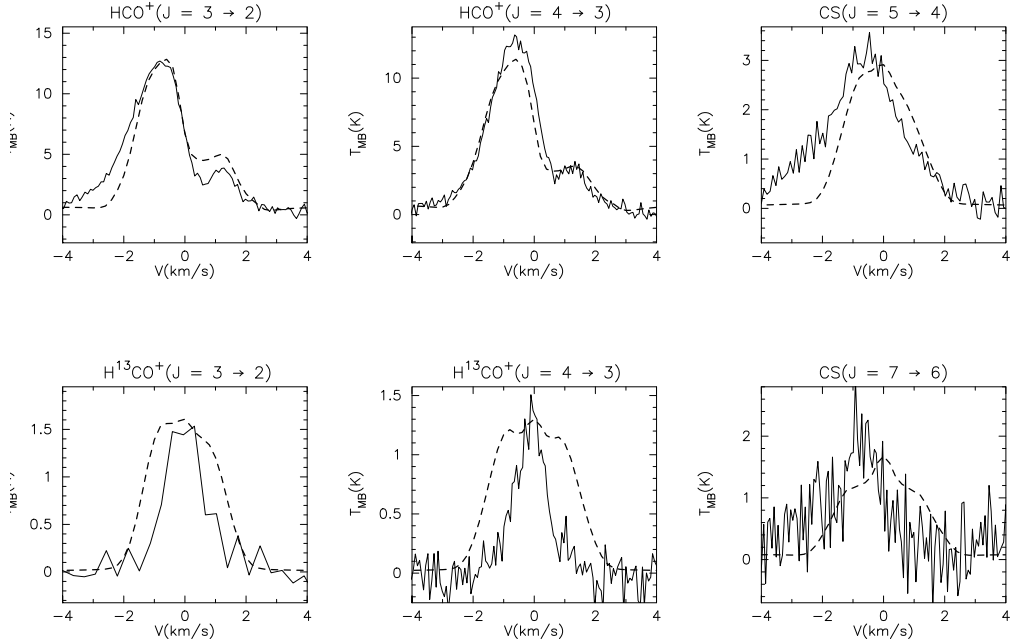}
\end{picture}
\caption{Model fits to the CS and \hco 
on-source spectra observed towards Serpens SMM4.}
\label{smm4spec2}
\end{figure*}

The mass of the central protostars can be approximately estimated 
from the infall velocity profile used in the model 
fit, applying the freefall equation $M_{\star}=rv^2/2G$ to the 
inner part of the profile. 
Using this relation we obtain an estimate of $\simeq 0.3$M$_{\odot}$. 
However, this must be regarded as somewhat uncertain.
The density and velocity profiles adopted in this model are 
inconsistent with the SIS model, 
despite individually having the SIS model forms. The infall radius 
used in the velocity profile is a 
factor of $\sim 7$ greater than the infall radius used in the 
density profile, indicating that
the infall is more well developed than a SIS model envelope with 
a similar density profile. 

Hydrodynamical (e.g. Foster \& Chevalier 1993) and magneto-hydrodynamical 
(e.g. Ciolek \& Mouschovias 1994) simulations of the evolution 
of dense cloud cores during the
pre-stellar phase predict that by the time the density 
profile approximates closely to 
a singular distribution, substantial infall velocities have already 
developed. Notwithstanding the
uncertainties in our model fit, our results 
support this picture of a non-static initial condition
for collapse.

\subsection{Serpens SMM4}

Figure~32 shows the model fits to the CS and \hco 
on-source spectra observed towards Serpens SMM4. 
The approach to the modelling was similar to that 
used for NGC1333--IRAS2, starting with
the \hcoft line, and then making adjustments to 
produce the optimal agreement with the
remaining lines. To produce a fit to the \hcoft 
line required the infall radius in the SIS model 
form of the velocity profile to be set well beyond 
the outer radius of the cloud, i.e. the entire 
protostellar envelope must be infalling. 

If the infall radius is set
inside the outer radius of the cloud, then absorption 
from the static gas beyond this radius
tends to produce an absorption dip close to the 
systemic velocity, in disagreement 
with the observed profile. As the SIS model velocity 
profile is inappropriate in this case, 
we used a simple $r^{-1/2}$ power law, which is the 
asymptotic form of the inner part
of the SIS model profile, to model the velocity profile instead. 

The parameters used in the model are given in Table~6.
The outer radius of the model envelope was taken as 
10\,000\,AU, and we used an innermost shell radius of 62\,AU.
The assumed distance to the source was 300\,pc. The total 
envelope mass implied by the model is $\sim$5.1\,M$_{\odot}$, roughly
consistent with the estimate of $\sim$3\,M$_{\odot}$ derived 
from the IRAS fluxes by Hurt \& Barsony (1996). 

\begin{table}
\begin{center}
\begin{tabular}{ll}\hline
Parameter & Value \\ \hline
$v_r(r) $ & $-1.95\left(\frac{r}{1000\,{\rm AU}}\right)^{-\frac{1}{2}} 
{\rm km\,s}^{-1}$\\
$n_{{\rm H}_2}(r<2000{\rm AU})$ & 
$1.48\times 10^{6}[ 0.35\left(\frac{r}{2000\,{\rm AU}}
\right)^{-\frac{3}{2}}$ \\
& \hspace*{0.25cm}
$+ 0.65\left(\frac{r}{2000\,{\rm AU}}\right)^{-0.64} ]\ {\rm cm}^{-3}$\\
$n_{{\rm H}_2}(r>2000\,{\rm AU})$ & 
$1.48\times 10^6 \left(\frac{r}{2000\,{\rm AU}}\right)^{-2}\ {\rm cm}^{-3}$\\
$T(r<3800\,{\rm AU})$ & $ 25\,\left(\frac{r}{3800{\rm AU}}\right)^{-0.4}\ 
{\rm K}$\\ 
$T(r>3800\,{\rm AU})$ & $25\,{\rm K}$\\ 
$\Delta v_{\rm tb}({\rm FWHM})$ & $0.80\,{\rm km\,s}^{-1}$\\ 
$X_{\rm HCO^+}$ & $3.5\times 10^{-9}$\\
$X_{\rm CS}$ & $3.5\times 10^{-9}$\\
$^{12}{\rm C}/^{13}{\rm C}$ & $90$ \\ \hline 
\end{tabular}
\caption{The parameters used in the model fit for Serpens~SMM4.}
\end{center}
\end{table}

As with NGC1333--IRAS2,
the infall velocities in the model fit are much greater than would be 
predicted by
the SIS model given the density profile. From the infall velocity 
profile we estimate the mass 
of the central protostellar system to be $M_{\star}=rv^2/2G=2.1$M$_{\odot}$. 
The temperature profile is normalised at
a considerably higher level than in the IRAS2 fit, which would 
suggest that SMM4 is considerably 
more luminous. 

\begin{figure*}
\setlength{\unitlength}{1mm}
\begin{picture}(60,190)
\includegraphics{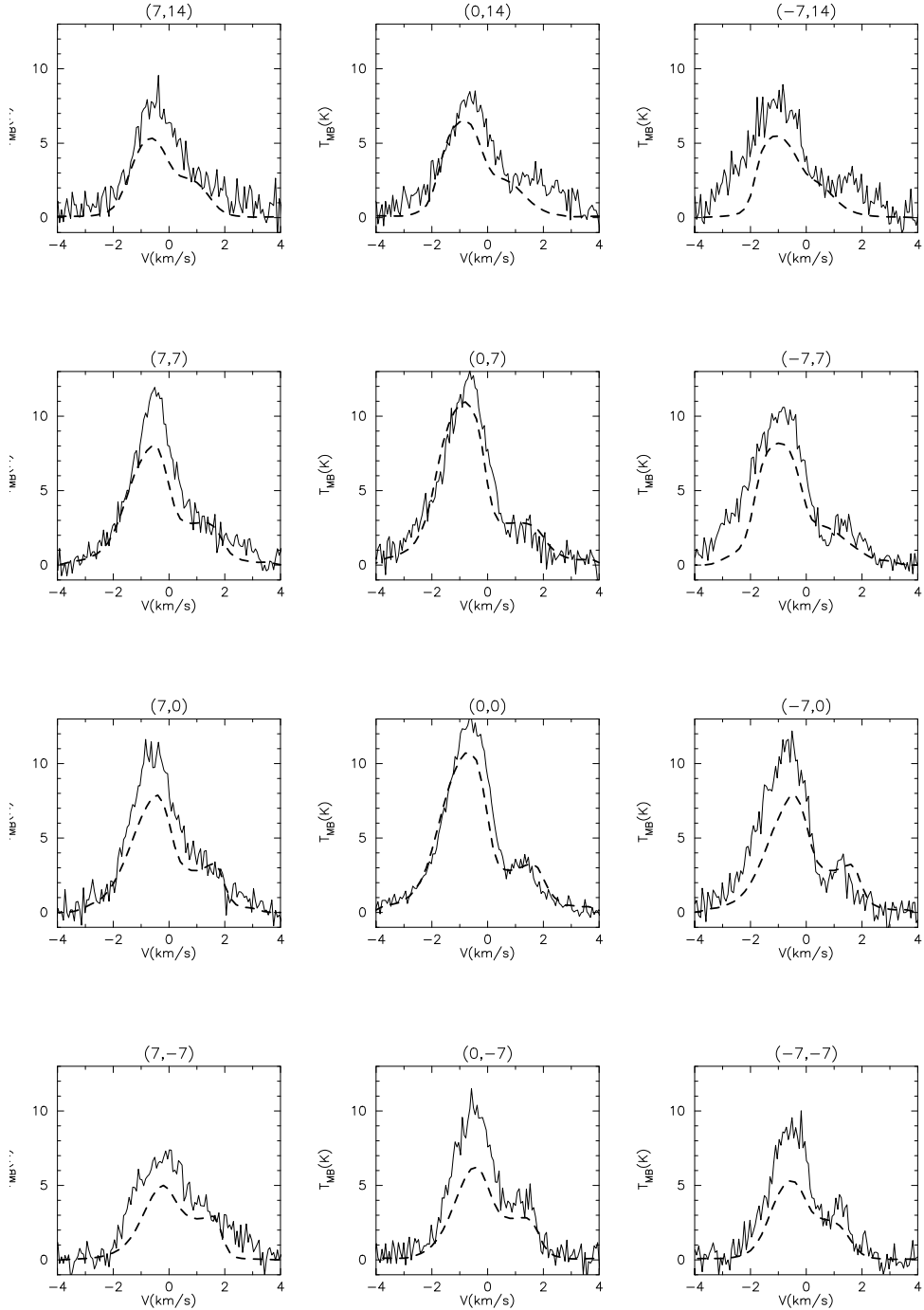}
\end{picture}
\caption{Comparison of the off-centre \hcoft 
line profiles observed towards Serpens SMM4 
(solid lines) with the model predictions (dashed lines). The
offsets (in arcsec) from the IRAS2 position 
are given in the title of each plot. In 
the model spectra we have included a solid body 
rotation of 28\,\kms\,pc$^{-1}$, about an 
axis passing through the centre of the cloud at a 
position angle of +40$^{\circ}$.
The centre of the cloud was assumed 
to lie at the offset (0,+3.5).}
\label{smm4specmap}
\end{figure*}

A high accretion luminosity for this object is 
also suggested by the large central 
protostellar mass and mass accretion rate ($\simeq 6\times 
10^{-5}$\,M$_{\odot}$\,yr$^{-1}$). 
Hurt \& Barsony (1996) estimated the luminosity of SMM4 as
9\,L$_{\odot}$, after 
deconvolving the HIRES IRAS images of the Serpens region to 
obtain the far-infrared fluxes. 
Jennings et al. (1987) estimated the luminosity of IRAS2 as 
17L$_{\odot}$, also using IRAS fluxes,
so there are clearly large uncertainties on these estimates. 
The parameters used in our model fits are inconsistent with 
these luminosities, and for 
our model to be correct, then either the 
SMM4 luminosity calculated from the IRAS data,
or our assumed protostellar radius,
would have to be a significant underestimate. 

The predicted \htco linewidths are considerably broader than the 
observations, as we found with NGC1333--IRAS2, although the peak 
intensities are in very good 
agreement. Again, we suggest this may be indicating deviations of 
the infall velocity 
profile below the $r^{-1/2}$ free-fall profile in the centre of 
the cloud. We found above
that the SMM4 envelope shows strong evidence for rotation 
($\Omega \sin i \simeq 28\,$\kms\,pc$^{-1}$),
and suggested that the flattening seen in the spatial distribution 
of integrated \hcoft emission
may be caused by centrifugal support of the outer envelope. Thus 
centrifugal braking may retard the infall velocity near the centre of 
the cloud. 

The calculated \csff and \csss line 
profiles match the observed line intensities reasonably well,
but the peak velocities of the calculated profiles are less red-shifted 
than the observations. 
The enhanced blue-shifted wings seen in both of the observed CS spectra 
strongly suggest
that outflow contributes to the emission in these lines. As discussed
above, CS is believed 
to be enhanced relative to \hco in bipolar outflows (Blake et al. 1995).

Figure~33 shows the fits to the off-centre \hcoft line profiles towards 
SMM4 from 
our best-fit model. We included a solid body rotation of 28\kms\,pc$^{-1}$ 
in the model
(assuming rotation axis in plane of sky), consistent
with the above centroid velocity gradient across SMM4. As the 
\hcoft spectra at the (0,0) and (0,+7) positions have approximately 
equal intensities,
we assumed that the centre of the cloud lies at the offset (0,+3.5). 
Although the model profiles
somewhat underestimate the observed lines, we are encouraged by 
the overall agreement of the trends in the observed lineshapes across 
the envelope. 
It is likely that a better fit to the line intensities
requires a shallower density profile than
predicted by the SIS model. 

In contrast to IRAS2 we find little evidence for
outflow emission in the off-centre line profiles, and essentially all of the 
high velocity emission can be accounted for by the infall. We consider these 
results to provide 
convincing evidence that both infall and rotation are taking place in the
the Serpens SMM4 envelope. The magnitude of the rotational velocity is large
enough for centrifugal effects to have a significant influence on the gas 
dynamics. 

\section{Summary}

In this paper
we have discussed the physical constraints which can be used to limit
the number and range of parameters needed to define the spherically symmetric 
models in the Stenholm radiative transfer code.
We have studied the dependence of the predicted line profiles
on a number of model parameters. Increasing the infall velocities in the 
envelope tends to 
produce stronger asymmetry between the blue and red-shifted peaks of 
double-peaked line profiles, 
without strongly affecting the separation between the peaks. {\em The value 
of the turbulent velocity
dispersion is the most significant factor in determining
the velocity separation between the peaks}. 
We carried out a preliminary investigation into the effect of adopting 
different
power law dependences of the turbulent velocity width on the gas density 
and found the
predicted line profiles to be extremely sensitive to the value of the 
power law exponent.

We have studied the effect on the predicted line profiles of super-imposing 
a solid-body rotation
onto the infall velocity field. The effect of the solid-body rotation on
the on-source double-peaked line profiles was to `smooth out' the 
double-peaked structure.
The centroid velocities and the line shapes of the off-centre profiles
were found to be significantly altered by the solid-body rotation.
On the red-shifted side of the rotational velocity gradient, a 
reversal of the asymmetry in the line profiles was seen.
We measured the centroid velocity gradient
predicted by our model for a solid-body 
rotating envelope, which was found to
slightly underestimate the actual rotational 
velocity gradient used in the model.

We selected the objects NGC1333--IRAS2 and Serpens~SMM4 for detailed 
modelling
using the Stenholm radiative transfer code, confining our search to 
the SIS model forms for 
the density and velocity profiles in the modelling, although we did 
not require
these profiles to use the same SIS model parameters in the models.
Our best model fits to both the NGC1333--IRAS2 and Serpens~SMM4 data 
used velocity profiles
which were more strongly developed than the SIS model would predict, 
given the fit to the density
profile. This is consistent with theoretical predictions that at the 
instant a central protostar is
formed, the gas in the envelope has already accelerated to substantial 
velocities
(e.g. Foster \& Chevalier 1993; JCH), which contrasts with the static 
initial state
assumed by the SIS model, and is one of the principal distinguishing 
features between
the SIS model and other collapse models.

Our fits to the \htco line profile observations of both objects 
slightly over-estimated the
profile widths. We could produce better fits to these lines by 
lowering the infall velocity
in the centre of the cloud, although this affected the 
fits to the main line profile slightly.
The infall velocity may therefore be retarded from the $r^{-1/2}$ 
free-fall relation towards the
centre of the cloud, possibly by centrifugal braking. 
In the case of IRAS2, the discrepancies between the model 
predictions and the observations
for the off-source profiles could not possibly be accommodated
within a plausible model of infall and rotation, and we 
attribute them to the outflow.

The model fits to the off-centre \hcoft line profiles of Serpens SMM4 
included the
solid-body rotation inferred from the \hcoft centroid velocity gradient 
in the calculation.
The predicted and observed profiles showed good agreement in the overall 
line shape although
the model slightly underestimated the line intensities. 
We thus
conclude that Serpens SMM4 is an infalling and rotating protostar.

\section*{Acknowledgments}

The authors would like to thank the staff of the Joint Astronomy Centre,
Hawaii (JACH), 
and in particular the JCMT Telescope Support Staff, for assistance
during the taking of the observations presented in this paper. We would also
like to thank Les Little for giving us a copy of the Kent Stenholm code.
The JACH is operated jointly by the UK PPARC, Canadian NSF and Netherlands
NWO.

\end{document}